%% file: neurips_2023.tex
\documentclass{article}

\PassOptionsToPackage{square,comma,numbers,sort&compress}{natbib}
\usepackage[preprint]{neurips_2023}

\usepackage{notoccite}

\usepackage[utf8]{inputenc} % allow utf-8 input
\usepackage[T1]{fontenc}    % use 8-bit T1 fonts
\usepackage{hyperref}       % hyperlinks
\usepackage{url}            % simple URL typesetting
\usepackage{booktabs}       % professional-quality tables
\usepackage{amsfonts}       % blackboard math symbols
\usepackage{nicefrac}       % compact symbols for 1/2, etc.
\usepackage{microtype}      % microtypography
\usepackage{xcolor}         % colors
\usepackage{caption}
\usepackage{subcaption}
\usepackage{graphicx}
\usepackage{amssymb}
\usepackage{import}
\usepackage{algorithm,algorithmicx,algpseudocode}
\usepackage{multirow}
\usepackage{tablefootnote}
\usepackage{pifont}% http://ctan.org/pkg/pifont
\newcommand{\cmark}{\ding{51}}%
\newcommand{\xmark}{\ding{55}}%

\usepackage{amsmath}
\usepackage{amsthm}

\title{Efficient Neural Music Generation}

\author{%
  Max W. Y. Lam, Qiao Tian, Tang Li, Zongyu Yin, Siyuan Feng, Ming Tu, Yuliang Ji,\\
  \textbf{Rui Xia, Mingbo Ma, Xuchen Song, Jitong Chen, Yuping Wang, Yuxuan Wang}\\
  Speech, Audio \& Music Intelligence (SAMI), ByteDance\\
}

\begin{document}

\maketitle

\begin{abstract}
Recent progress in music generation has been remarkably advanced by the state-of-the-art MusicLM, which comprises a hierarchy of three LMs, respectively, for semantic, coarse acoustic, and fine acoustic modelings. Yet, sampling with the MusicLM requires processing through these LMs one by one to obtain the fine-grained acoustic tokens, making it computationally expensive and prohibitive for a real-time generation. Efficient music generation with a quality on par with MusicLM remains a significant challenge.
In this paper, we present \textbf{M}e\textbf{L}o\textbf{D}y (\textbf{M} for music; \textbf{L} for LM; \textbf{D} for diffusion), an LM-guided diffusion model that generates music audios of state-of-the-art quality meanwhile reducing 95.7\% or 99.6\% forward passes in MusicLM, respectively, for sampling 10s or 30s music. MeLoDy inherits the highest-level LM from MusicLM for semantic modeling, and applies a novel dual-path diffusion (DPD) model and an audio VAE-GAN to efficiently decode the conditioning semantic tokens into waveform. DPD is proposed to simultaneously model the coarse and fine acoustics by incorporating the semantic information into segments of latents effectively via cross-attention at each denoising step. Our experimental results suggest the superiority of MeLoDy, not only in its practical advantages on sampling speed and infinitely continuable generation, but also in its state-of-the-art musicality, audio quality, and text correlation.\\
Our samples are available at \url{https://Efficient-MeLoDy.github.io/}.
\end{abstract}

\iffalse
Something to note (if any):
- 
\fi

\section{Introduction}

Music is an art composed of harmony, melody, and rhythm that permeates every aspect of human life. With the blossoming of deep generative models \cite{ho2020denoising, rombach2022high, borsos2022audiolm}, music generation has drawn much attention in recent years \cite{dhariwal2020jukebox, agostinelli2023musiclm, huang2023noise2music}. As a prominent class of generative models, language models (LMs) \cite{vaswani2017attention, devlin2018bert} showed extraordinary modeling capability in modeling complex relationships across long-term contexts \cite{chen2021evaluating, thoppilan2022lamda, ouyang2022training}. In light of this, AudioLM \cite{borsos2022audiolm} and many follow-up works \cite{kreuk2022audiogen, agostinelli2023musiclm, wang2023neural, kharitonov2023speak} successfully applied LMs to audio synthesis. Concurrent to the LM-based approaches, diffusion probabilistic models (DPMs) \cite{jascha2015, ho2020denoising, kingma2021variational}, as another competitive class of generative models \cite{dhariwal2021diffusion, rombach2022high}, have also demonstrated exceptional abilities in synthesizing speech \cite{huang2022fastdiff, kim2022guided, shen2023naturalspeech}, sounds \cite{liu2023audioldm, huang2023make} and music \cite{huang2023noise2music, schneider2023mo}.

However, generating music from free-form text remains challenging as the permissible music descriptions can be very diverse and relate to any of the genres, instruments, tempo, scenarios, or even some subjective feelings. Conventional text-to-music generation models are listed in Table~\ref{tab:previous-works}, where both MusicLM \cite{agostinelli2023musiclm} and Noise2Music \cite{huang2023noise2music} were trained on large-scale music datasets and demonstrated the state-of-the-art (SOTA) generative performances with high fidelity and adherence to various aspects of text prompts. Yet, the success of these two methods comes with large computational costs, which would be a serious impediment to their practicalities. In comparison, Mo\^{u}sai \cite{schneider2023mo} building upon DPMs made efficient samplings of high-quality music possible. Nevertheless, the number of their demonstrated cases was comparatively small and showed limited in-sample dynamics. Aiming for a feasible music creation tool, a high efficiency of the generative model is essential since it facilitates interactive creation with human feedback being taken into account as in \cite{holz2022midjourney}.

While LMs and DPMs both showed promising results, we believe the relevant question is not whether one should be preferred over another but whether we can leverage both approaches with respect to their individual advantages, e.g., \cite{xiao2021tackling}. After analyzing the success of MusicLM, we leverage the highest-level LM in MusicLM, termed as \textit{semantic LM}, to model the semantic structure of music, determining the overall arrangement of melody, rhythm, dynamics, timbre, and tempo. Conditional on this semantic LM, we exploit the non-autoregressive nature of DPMs to model the acoustics efficiently and effectively with the help of a successful sampling acceleration technique \cite{song2020denoising}. All in all, in this paper, we introduce several novelties that constitute our main contributions:
% Distinctively real-time \cite{}, instead of jointly modeling semantic and acoustic information with LMs, we propose to use a non-autoregressive approach for acoustic modeling.

\begin{enumerate}
    \item We present \textbf{M}e\textbf{L}o\textbf{D}y (\textbf{M} for music; \textbf{L} for LM; \textbf{D} for diffusion), an LM-guided diffusion model that generates music of competitive quality while reducing 95.7\% and 99.6\% iterations of MusicLM to sample 10s and 30s music, being faster than real-time on a V100 GPU.
    \item We propose the novel dual-path diffusion (DPD) models to efficiently model coarse and fine acoustic information simultaneously with a particular semantic conditioning strategy. 
    \item We design an effective sampling scheme for DPD, which improves the generation quality over the previous sampling method in \cite{schneider2023mo} proposed for this class of LDMs.
    \item We reveal a successful audio VAE-GAN that effectively learns continuous latent representations, and is capable of synthesizing audios of competitive quality together with DPD.
\end{enumerate}

\begin{table}
  \caption{A comparison of MeLoDy with conventional text-to-music generation models in the literature. We use \textbf{AC} to denote whether audio continuation is supported, \textbf{FR} to denote whether the sampling is faster than real-time on a V100 GPU, \textbf{VT} to denote whether the model has been tested and demonstrated using various types of text prompts including instruments, genres, and long-form rich descriptions, and \textbf{MP} to denote whether the evaluation was done by music producers.}
  \label{tab:previous-works}
  \centering
  \begin{tabular}{lccccccc}
    \toprule
    % \multicolumn{2}{c}{Part}                   \\
    % \cmidrule(r){1-2}
    \textbf{Model} & \textbf{Prompts} & \textbf{Training Data} & \textbf{AC} & \textbf{FR} & \textbf{VT} & \textbf{MP}\\
    \midrule
    % Jukebox \cite{dhariwal2020jukebox} & Text (lyrics) & 1.2M songs & \cmark &\xmark& \xmark & \xmark \\
    % Riffusion \cite{forsgrenriffusion} & Text & - & \xmark & \cmark & - \\
    Moûsai \cite{schneider2023mo} & Text & 2.5k hours of music & \cmark &\cmark& \xmark& \xmark \\
    MusicLM \cite{agostinelli2023musiclm} & Text, Melody & 280k hours of  music & \cmark &\xmark & \cmark& \xmark\\
    Noise2Music \cite{huang2023noise2music}  & Text & 340k hours of music & \xmark &\xmark& \cmark& \xmark\\
    \midrule
    \textbf{MeLoDy} (Ours) & Text, Audio & 257k hours of music\tablefootnote{We focus on non-vocal music data by using an audio classifier \cite{kong2020panns} to filter out in-house music data with vocals. Noticeably, generating vocals and instrumental music simultaneously in one model is defective even in the SOTA works \cite{agostinelli2023musiclm, huang2023noise2music} because of the unnaturally sound vocals. While this work aims for generating production-level music, we improve the fidelity by reducing the tendency of generating vocals.} & \cmark &\cmark & \cmark & \cmark\\
    \bottomrule
  \end{tabular}
  \vspace{-0.68em}
\end{table}

\section{Related Work}

\paragraph{Audio Generation}
Apart from the generation models shown in Table~\ref{tab:previous-works}, there are also music generation models \cite{caillon2021rave, pasini2022musika} that can generate high-quality music samples at high speed, yet they cannot accept free-form text conditions and can only be trained to specialize in single-genre music, e.g., techno music in \cite{pasini2022musika}. There also are some successful music generators in the industry, e.g. Mubert \cite{mubert} and Riffusion \cite{riffusion}, yet, as analyzed in \cite{agostinelli2023musiclm}, they struggled to compete with MusicLM in handling free-form text prompts. In a more general scope of audio synthesis, some promising text-to-audio synthesizers \cite{kreuk2022audiogen, liu2023audioldm, huang2023make} trained with AudioSet \cite{gemmeke2017audio} also demonstrated to be able to generate music from free-form text, but the musicality is limited. AudioLM \cite{borsos2022audiolm} unconditionally continued piano audios with promising fidelity. Parallel to this work, SoundStorm \cite{borsos2023soundstorm} exceedingly accelerated the AudioLM with a non-autoregressive decoding scheme \cite{chang2022maskgit}, such that the acoustic LM can be decoded in 27 forward passes. In comparison, neglecting the individual cost of networks, MeLoDy takes 5 to 20 forward passes to generate acoustics of high fidelity, as discussed in Section \ref{sec:exps}.

\paragraph{Network Architecture}
The architecture designed for our proposed DPD was inspired by the dual-path networks used in the context of audio separation, where \citet{luo2020dual} initiated the idea of segmentation-based dual-path processing, and triggered a number of follow-up works achieving the state-of-the-art results \cite{chen2020dual, lam2021effective, lam2021sandglasset, subakan2021attention, zhao2023mossformer}. Noticing that the objective in diffusion models indeed can be viewed as a special case of source separation, this kind of dual-path architecture effectually provides us a basis for simultaneous coarse-and-fine acoustic modeling.

\section{Background on Audio Language Modeling}

This section provides the preliminaries that serve as the basis for our model. In particular, we briefly describe the audio language modeling framework used in MusicLM.

\subsection{Audio Language Modeling with MusicLM}
MusicLM \cite{agostinelli2023musiclm} mainly follows the audio language modeling framework presented in AudioLM \cite{borsos2022audiolm}, where audio synthesis is viewed as a language modeling task over a hierarchy of coarse-to-fine audio
tokens. In AudioLM, there are two kinds of tokenization for representing different scopes of audio:
\begin{itemize}
    \item \textbf{Semantic Tokenization}: K-means over representations from SSL, e.g., w2v-BERT \cite{chung2021w2v};
    \item \textbf{Acoustic Tokenization}: Neural audio codec, e.g., SoundStream \cite{zeghidour2021soundstream}.
\end{itemize}
To better handle the hierarchical structure of the acoustic tokens, AudioLM further separates the modeling of acoustic tokens into coarse and fine stages. In total, AudioLM defines three LM tasks: (1) semantic modeling, (2) coarse acoustic modeling, and (3) fine acoustic modeling.

We generally define the sequence of conditioning tokens as $\mathbf{c}_{1:T_\text{cnd}}:=[\mathbf{c}_1, \ldots, \mathbf{c}_{T_\text{cnd}}]$ and the sequence of target tokens as $\mathbf{u}_{1:T_\text{tgt}}:=[\mathbf{u}_1, \ldots, \mathbf{u}_{T_\text{tgt}}]$. In each modeling task, a Transformer-decoder language model parameterized by $\theta$ is tasked to solve the following autoregressive modeling problem:
\begin{align}
    p_{\theta}(\mathbf{u}_{1:T_\text{tgt}}|\mathbf{c}_{1:T_\text{cnd}})=\prod_{j=1}^{T_\text{tgt}} p_\theta(\mathbf{u}_j|[\mathbf{c}_1, \ldots, \mathbf{c}_{T_\text{cnd}}, \mathbf{u}_1, \ldots, \mathbf{u}_{j-1}]),
    \label{eq:autoregressive-modeling}
\end{align}
where the conditioning tokens are concatenated to the target tokens as prefixes. In AudioLM, semantic modeling takes no condition; coarse acoustic modeling takes the semantic tokens as conditions; fine acoustic modeling takes the coarse acoustic tokens as conditions. The three corresponding LMs can be trained in parallel with the ground-truth tokens, but need to be sampled sequentially for inference.

\subsubsection{Joint Tokenization of Music and Text with MuLan and RVQ}
To maintain the merit of audio-only training, MusicLM relies on MuLan \cite{huang2022mulan}, which is a two-tower, joint audio-text embedding model that can be individually trained with large-scale music data and weakly-associated, free-form text annotations. The MuLan model is pre-trained to project the music audio and its corresponding text description into the same embedding space such that the associated embeddings can be close to each other. In MusicLM, the MuLan embeddings of music and text are tokenized using a separately learned residual vector quantization (RVQ) \cite{zeghidour2021soundstream}.

Different from AudioLM, MusicLM employs the MuLan tokens as the additional prefixing tokens, as in Eq. (\ref{eq:autoregressive-modeling}), for the semantic modeling and the coarse acoustic modeling. During training, the audio is first fed to the MuLan music tower to obtain the music embedding. Then, an RVQ is applied to the music embedding, resulting in the ground-truth MuLan tokens for conditioning the semantic LM and the coarse acoustic LM. To generate music from a text prompt, the text embedding obtained from the MuLan text tower is passed to the same RVQ and is discretized into the inference-time MuLan tokens. Based on the prefixing MuLan tokens, the semantic tokens, coarse acoustic tokens, and fine acoustic tokens are subsequently computed to generate high-fidelity music audio adhering to the text prompt.

\begin{figure}
     \centering
     \includegraphics[width=\textwidth]{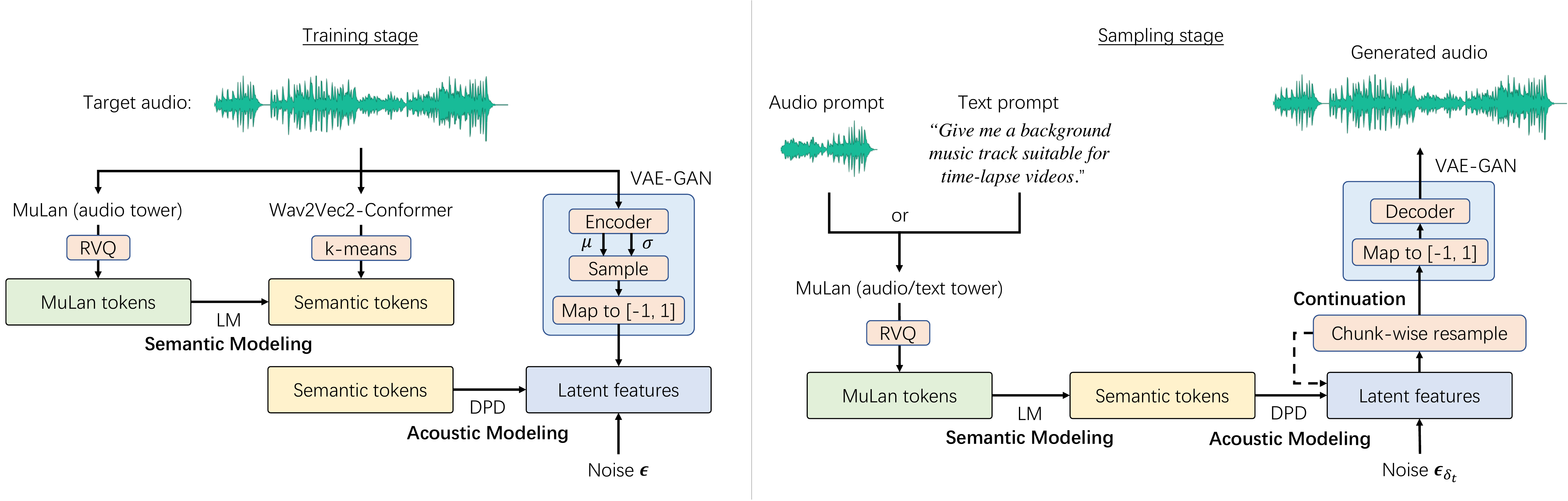}
     \caption{The training and sampling pipelines of MeLoDy}
     \label{fig:overall}
  \vspace{-0.68em}
\end{figure}

\section{Model Description}
\label{sec:model}

The overall training and sampling pipelines of MeLoDy are shown in Figure~\ref{fig:overall}, where, we have three modules for representation learning: (1) MuLan, (2) Wav2Vec2-Conformer, and (3) audio VAE, and two generative models: a language model (LM) and a dual-path diffusion (DPD) model, respectively, for semantic modeling and acoustic modeling. In the same spirit as MusicLM, we leverage LM to model the semantic structure of music for its promising capability of modeling complex relationships across long-term contexts \cite{chen2021evaluating, thoppilan2022lamda, ouyang2022training}. Similar to MusicLM, we pre-train a MuLan model to obtain the conditioning tokens. For semantic tokenization, we opt to use the Wav2Vec2-Conformer model, which follows the same architecture as Wav2Vec2 \cite{baevski2020wav2vec} but employs the Conformer blocks \cite{gulati2020conformer} in place of the Transformer blocks. The remainder of this section presents our newly proposed DPD model and the audio VAE-GAN used for DPD model, while other modules overlapped with MusicLM are described in Appendix B regarding the training and implementation details.

\subsection{Dual-Dath Diffusion: Angle-Parameterized Continuous-Time Latent Diffusion Models}
The proposed dual-path diffusion (DPD) model is a variant of diffusion probabilistic models (DPMs) \cite{jascha2015, ho2020denoising, nichol2021improved} in continuous-time \cite{song2020score, kingma2021variational, salimans2022progressive, karras2022elucidating}. Instead of directly operating on the raw data $\mathbf{x}\sim p_\text{data}(\mathbf{x})$, with reference to the latent diffusion models (LDMs) \cite{rombach2022high}, we consider a low-dimensional latent representation $\mathbf{z}=\mathcal{E}_\phi(\mathbf{x})$, where $\phi$ is a pre-trained autoencoder that enables reconstruction of the raw data from the latent: $\mathbf{x}\approx\mathcal{D}_\phi(\mathbf{z})$. Here, we use $\mathcal{E}_\phi$ to denote the encoder, and $\mathcal{D}_\phi$ to denote the decoder. By working on a low-dimensional latent space, the computational burden of DPMs can be significantly relieved \cite{rombach2022high}. We present our audio autoencoder in Section \ref{sec:kl-vae}, which is tailored for DPMs and performed the stablest in our experiments.

In DPD, we consider a Gaussian diffusion process $\mathbf{z}_t$ that is fully specified by two strictly positive scalar-valued, continuously differentiable functions $\alpha_t, \sigma_t$ \cite{kingma2021variational}: $q(\mathbf{z}_t|\mathbf{z})=\mathcal{N}(\mathbf{z}_t; \alpha_t \mathbf{z}, \sigma_t^2 \mathbf{I})$ for any $t \in [0, 1]$.
% By marginalization, any finite collection with $0\leq s < t \leq 1$ satisfies a nice Markovian structure: $q(\mathbf{z}_t|\mathbf{z}_s)=\mathcal{N}(\mathbf{z}_t; (\alpha_t/\alpha_s) \mathbf{z}_s, (\sigma_t^2-(\alpha_t/\alpha_s)\sigma_s^2) \mathbf{I})$.
In the light of \cite{salimans2022progressive}, we define $\alpha_t:=\cos(\pi t/2)$ and $\sigma_t:=\sin(\pi t/2)$ to benefit from some nice trigonometric properties, i.e.,  $\sigma_t=\sqrt{1-\alpha_t^2}$ (a.k.a. variance-preserving \cite{kingma2021variational}). By this definition, $\mathbf{z}_t$ can be elegantly re-parameterized in terms of angles $\delta$:
\begin{align}
\mathbf{z}_{\delta}=\cos(\delta)\mathbf{z}+\sin(\delta)\boldsymbol\epsilon\,\,\,\,\,\text{for any }\,\,\delta\in [0, \pi/2],\,\,\,\,\,\boldsymbol\epsilon\sim\mathcal{N}(\mathbf{0}, \mathbf{I}).
\end{align}
Note that $\mathbf{z}_\delta$ gets noisier as $\delta$ increases from $0$ to $\pi/2$, which defines the forward diffusion process.

To generate samples, we use a $\theta$-parameterized variational model $p_\theta(\mathbf{z}_{\delta-\omega}|\mathbf{z}_{\delta})$ to invert the diffusion process by enabling running backward in angle with $0<\omega\leq\delta$. Based on this model, we can sample $\mathbf{z}$ from $\mathbf{z}_{\pi/2}\sim\mathcal{N}(\mathbf{0}, \mathbf{I})$ with $T$ sampling steps, by discretizing $\pi/2$ into $T$ segments as follows:
\begin{align}
p_\theta(\mathbf{z}|\mathbf{z}_{\pi/2})=\int_{\mathbf{z}_{\delta_{1:T-1}}} \prod_{t=1}^{T}p_\theta(\mathbf{z}_{\delta_{t}-\omega_t}|\mathbf{z}_{\delta_t})\,d\mathbf{z}_{\delta_{1:T-1}},\,\,\,\,\,\delta_{t}=\begin{cases}
    \frac{\pi}{2}-\sum_{i=t+1}^T\omega_i, & 1 \leq t < T;\\
    \frac{\pi}{2}, & t = T,
\end{cases}
\label{eq:delta-t-1}
\end{align}
where the \textit{angle schedule}, denoted by $\omega_1, \dots, \omega_T$, satisfies $\sum_{t=1}^{T}\omega_t=\pi/2$. \citet{schneider2023mo} proposed a uniform angle schedule: $\omega_t=\frac{\pi}{2T}$ for all $t$. As revealed in previous scheduling methods \cite{chen2020wavegrad,lam2022bddm} for DPMs, taking larger steps at the beginning of the sampling followed by smaller steps could improve the quality of samples. Following this strategy, we design a new linear angle schedule, which empirically gives more stable and higher-quality results, and is written as
\begin{align}
\omega_t=\frac{\pi}{6T}+\frac{2\pi t}{3T(T+1)}.
\label{eq:angle-schedule}
\end{align}
We extensively compare this linear angle schedule with the uniform one in \cite{schneider2023mo} in Appendix D.

\begin{figure}
     \centering
    \includegraphics[width=0.55\textwidth]{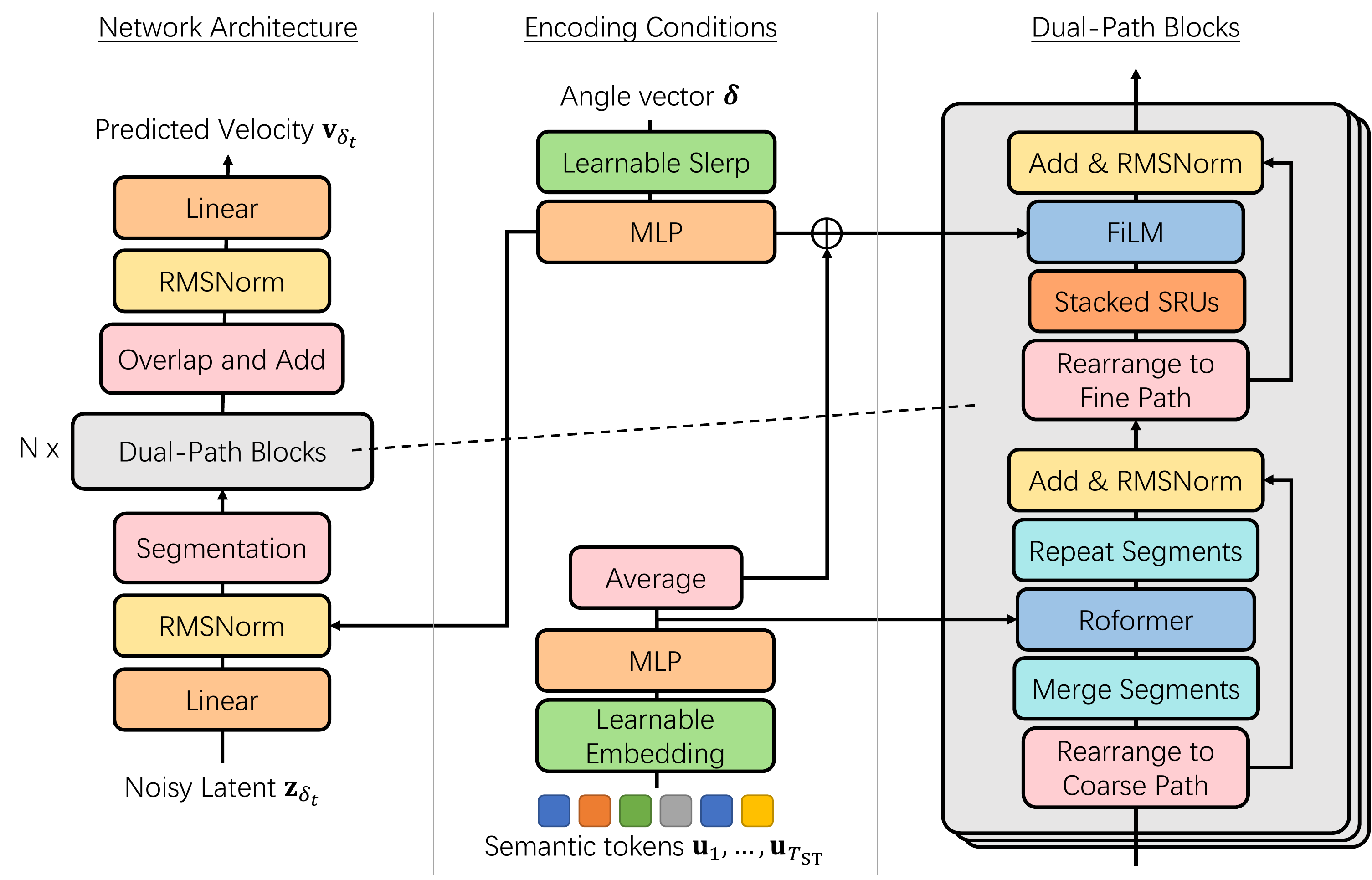}
     \caption{The proposed dual-path diffusion (DPD) model}
        \label{fig:dualpath-diff}
  \vspace{-0.68em}
\end{figure}
\subsubsection{Multi-Chunk Velocity Prediction for Long-Context Generation}
For model training, similar to the setting in \cite{schneider2023mo} for long-context generation, the neural network is tasked to predict a multi-chunk target $\mathbf{v}_\text{tgt}$ that comprises $M$ chunks of velocities, each having a different noise scale. Formally speaking, given that $\mathbf{z}, \mathbf{z}_\delta, \boldsymbol\epsilon \in\mathbb{R}^{L\times D}$ with $L$ representing the length of audio latents and $D$ representing the latent dimensions, we define $\mathbf{v}_\text{tgt}:=\mathbf{v}_1\oplus\cdots\oplus \mathbf{v}_M$, where
\begin{align}
\mathbf{v}_m:=\cos({\delta_m})\boldsymbol\epsilon[L_{m-1}:L_{m}, :]-\sin({\delta_m})\mathbf{z}[L_{m-1}:L_{m}, :],\,\,\,\,\,L_m:=\left\lfloor \frac{mL}{M} \right\rfloor.
\end{align}
Here, we use the NumPy slicing syntax ($0$ as the first index) to locate the $m$-th chunk, and we draw $\delta_m\sim\text{Uniform}[0, \pi/2]$ for each chunk at each training step to determine the noise scale. To learn $\theta$, we use the mean squared error (MSE) loss in \cite{ho2020denoising, salimans2022progressive}:
\begin{align}
    \mathcal{L}_\text{diff}&:=\mathbb{E}_{\mathbf{z}, {\boldsymbol\epsilon}, \delta_1, \ldots, \delta_M}\left[ \left\lVert\mathbf{v}_\text{tgt}-\hat{\mathbf{v}}_\theta(\mathbf{z}_\text{noisy};\mathbf{c})\right\rVert^2_2\right],\\\mathbf{z}_\text{noisy}&:=\cos({\delta_m})\mathbf{z}[L_{m-1}:L_{m}, :]+\sin({\delta_m})\boldsymbol\epsilon[L_{m-1}:L_{m}, :],
    \label{eq:diff-loss}
\end{align}
where $\mathbf{c}$ generally denotes the collection of conditions used for the velocity prediction.
% For example, in an unconditional generation, the diffusion index is typically taken as the condition to assist the denoising task, i.e., $\mathbf{c}=\{t\}$ \cite{ho2020denoising, kingma2021variational}.
% In text-to-music generation, Noise2Music \cite{huang2023noise2music} and Mo\^{u}ai \cite{schneider2023mo} took the text embeddings from the T5 encoder \cite{raffel2020exploring} for conditioning.
In MeLoDy, as illustrated in Figure~\ref{fig:overall}, we propose to use the semantic tokens $\mathbf{u}_{1}, \ldots, \mathbf{u}_{T_\text{ST}}$, which are obtained from the SSL model during training and generated by the LM at inference time, to condition the DPD model. In our experiments, we find that the stability of generation can be significantly improved if we use token-based discrete conditions to control the semantics of the music and let the diffusion model learn the embedding vector for each token itself. Additionally, to assist the multi-chunk prediction, we append an angle vector to the condition that represents the angles drawn in the $M$ chunks:
\begin{align}
    \mathbf{c}:=\left\{\mathbf{u}_{1}, \ldots, \mathbf{u}_{T_\text{ST}}, \boldsymbol\delta\right\}, \,\,\,\,\,\boldsymbol\delta:={[\delta_1]}_{r=1}^{L_1}\oplus\cdots\oplus{[\delta_M]}_{r=1}^{L_M}\in\mathbb{R}^{L}
    \label{eq:condition}
\end{align}
where ${[a]}_{r=1}^{B}$ denotes the operation of repeating a scalar $a$ for $B$ times to make a $B$-length vector.
Suppose we have a well-trained velocity model, for sampling, we apply the trigonometric identities to the DDIM sampling algorithm \cite{song2020denoising} (see Appendix A) and obtain a simplified update rule:
\begin{align}
    \mathbf{z}_{{\delta_t}-\omega_t}=\cos(\omega_t)\mathbf{z}_{\delta_t}-\sin(\omega_t)\hat{\mathbf{v}}_\theta(\mathbf{z}_{\delta_t};\mathbf{c}),
    \label{eq:update-rule}
\end{align}
by which, using the angle schedule in Eq. (\ref{eq:angle-schedule}) and running from $t=T$ to $t=1$, we get a sample of $\mathbf{z}$.

\subsubsection{Dual-Path Modeling for Efficient and Effective Velocity Prediction}
Next, we present how $\hat{\mathbf{v}}_\theta$ takes in the noisy latent and the conditions and efficiently incorporates the semantic tokens into the coarse processing path for effective velocity prediction. As a highlight of this work, we modify the dual-path technique borrowed from audio separation \cite{luo2020dual, lam2021effective, lam2021sandglasset}, and propose a novel architecture for efficient, simultaneous coarse and fine acoustic modeling, as shown in Figure~\ref{fig:dualpath-diff}. This architecture comprises several critical modules, which we present one by one below.

To begin with, we describe how the conditions are processed in DPD (the middle part in Figure~\ref{fig:dualpath-diff}):

\paragraph{Encoding Angle Vector} First, we encode ${\boldsymbol{\delta}}\in\mathbb{R}^{L}$, which records the frame-level noise scales of latents. Instead of using the classical positional encoding \cite{ho2020denoising}, we use a Slerp-alike spherical interpolation \cite{shoemake1985animating} to two learnable vectors $\mathbf{e}_\text{start}, \mathbf{e}_\text{end}\in\mathbb{R}^{256}$ based on broadcast multiplications $\otimes$:
\begin{align}
\mathbf{E}_{\boldsymbol{\delta}}:=\text{MLP}\left(\sin({\boldsymbol{\delta}})\otimes \mathbf{e}_\text{start} + \sin({\boldsymbol{\delta}})\otimes \mathbf{e}_\text{end}\right)\in\mathbb{R}^{L\times D_\text{hid}},
\end{align}
where $\text{MLP}(\mathbf{x}):=\text{RMSNorm}(\mathbf{W}_2\text{GELU}(\mathbf{x}\mathbf{W}_1+\mathbf{b}_1)+\mathbf{b}_2)$ projects an arbitrary input $\mathbf{x}\in\mathbb{R}^{D_\text{in}}$ to $\mathbb{R}^{D_\text{hid}}$ using RMSNorm \cite{zhang2019root} and GELU activation \cite{hendrycks2016gaussian}. Here, $D_\text{hid}$ is hidden dimension defined for the model, and $\mathbf{W}_1\in\mathbb{R}^{D_\text{in}\times D_\text{hid}}$, $\mathbf{W}_2\in\mathbb{R}^{D_\text{hid}\times D_\text{hid}}$, $\mathbf{b}_1,\mathbf{b}_2\in\mathbb{R}^{D_\text{hid}}$ are the learnable parameters.

\paragraph{Encoding Semantic Tokens} The remaining conditions are the discrete tokens representing semantic information $\mathbf{u}_{1}, \ldots, \mathbf{u}_{T_\text{ST}}$. Following the typical approach for embedding natural languages \cite{devlin2018bert}, we directly use a lookup table of vectors to map any token $\mathbf{u}_t\in \{1, \ldots, V_\text{ST}\}$ into a real-valued vector $E({\mathbf{u}_t})\in\mathbb{R}^{D_\text{hid}}$, where $V_\text{ST}$ denotes the vocabulary size of the semantic tokens, i.e., the number of clusters in k-means for Wav2Vec2-Conformer. By stacking the vectors along the time axis and applying an $\text{MLP}$ block, we obtain $\mathbf{E}_\text{ST}:=\text{MLP}\left(\left[E(\mathbf{u}_{1}), \ldots, E(\mathbf{u}_{T_\text{ST}})\right]\right)\in\mathbb{R}^{T_\text{ST}\times D_\text{hid}}$.

Next, we show how the network input (i.e., $\mathbf{z}_\text{noisy}$ at training time, or $\mathbf{z}_{\delta_t}$ at inference time) is processed given the condition embeddings. We use $\mathbf{z}_\text{noisy}$ as input for our explanation below, since $\mathbf{z}_{\delta_t}$ is only its special case with all chunks having the same noise scale. The input $\mathbf{z}_\text{noisy}$ is first linearly transformed and added up with the angle embedding of the same shape:
$
\mathbf{H}:=\text{RMSNorm}\left(\mathbf{z}_\text{noisy}\mathbf{W}_\text{in}+\mathbf{E}_{\boldsymbol{\delta}}\right),
$
where $\mathbf{W}_\text{in}\in\mathbb{R}^{D\times D_\text{hid}}$ is learnable. We then perform segmentation for dual-path processing.

\begin{figure}
     \centering
     \begin{subfigure}[b]{0.49\textwidth}
         \centering
         \includegraphics[width=\textwidth]{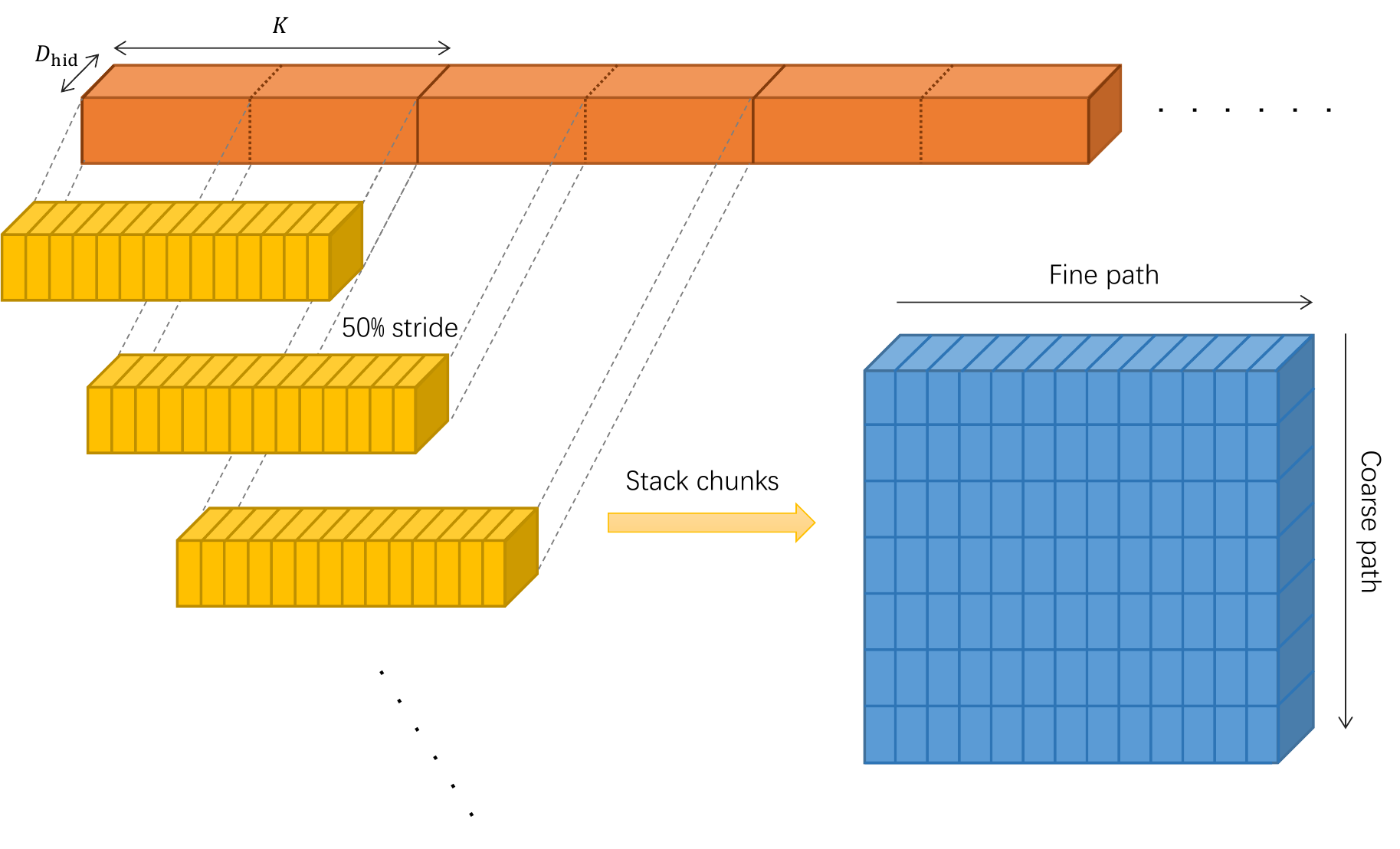}
         \caption{Segmentation}
         \label{fig:segmentation}
     \end{subfigure}
     \hfill
     \begin{subfigure}[b]{0.49\textwidth}
         \centering
         \includegraphics[width=\textwidth]{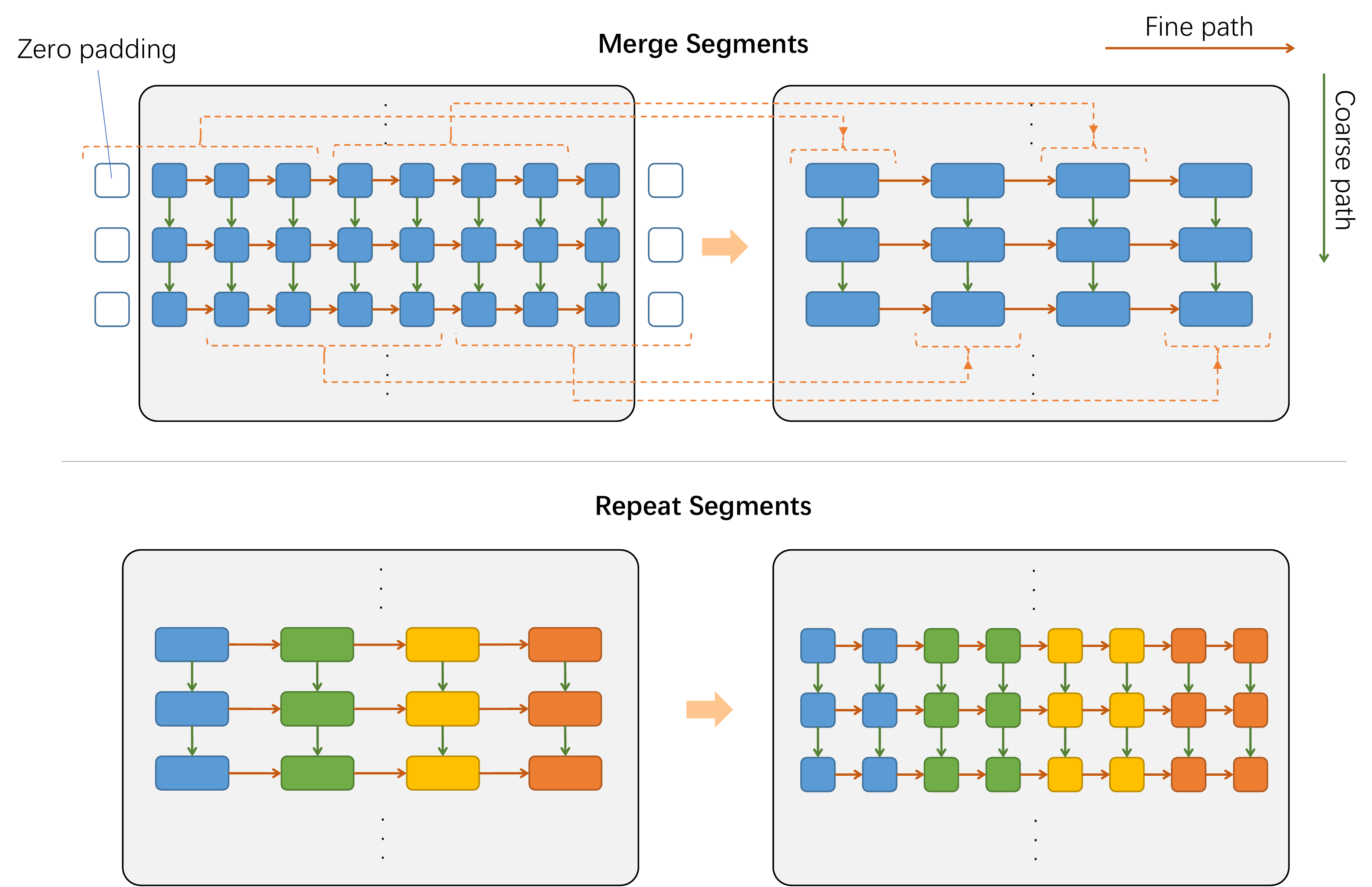}
         \caption{Merge Segments and Repeat Segments}
         \label{fig:chunk-ops}
     \end{subfigure}
     \caption{Diagrams for visually understanding the operations over the 3-D segments}
  \vspace{-0.68em}
\end{figure}

\paragraph{Segmentation}
As shown in Figure~\ref{fig:segmentation}, the segmentation module divides a 2-D input into $S$ half-overlapping segments each of length $K$, represented by a 3-D tensor $\mathbb{H}:=\left[\mathbf{0}, \mathbf{H}_{1}, \ldots, \mathbf{H}_{S}, \mathbf{0}\right]\in\mathbb{R}^{S\times K\times D_\text{hid}}$, where
$
\mathbf{H}_{s}:=\mathbf{H}\left[\frac{(s-1)K}{2}:\frac{(s-1)K}{2}+K, :\right],
$
and $\mathbb{H}$ is zero-padded such that we have $S=\left\lceil\frac{2L}{K}\right\rceil+1$. With a segment size $K\approx\sqrt{L}$, the length for sequence processing becomes sub-linear ($\mathcal{O}(\sqrt{L})$) as opposed to tackling the whole sequence ($\mathcal{O}(L)$). This greatly reduces the difficulty of learning a very long sequence and permits MeLoDy to use higher-frequency latents.

\paragraph{Dual-Path Blocks} After the segmentation, we obtain a 3-D tensor input for $N$ dual-path blocks, each block exhibits an architecture shown on the rightmost of Figure~\ref{fig:dualpath-diff}. The input to the $i$-th dual-path block is denoted as $\mathbb{H}^{(i)}$, and we have $\mathbb{H}^{(1)}:=\mathbb{H}$. Each block contains two stages corresponding to coarse-path (i.e., inter-segment) and fine-path (i.e., intra-segment) processing, respectively. Similar to the observations in \cite{lam2021effective, lam2021sandglasset}, we find it superior to use an attention-based network for coarse-path processing and to use a bi-directional RNN for fine-path processing. The goal of fine acoustic modeling is to better reconstruct the fine details from the roughly determined audio structure \cite{borsos2022audiolm}. At a finer scope, only the nearby elements matter and contain the most information needed for refinement, as supported by the modeling perspectives in neural vocoding \cite{oord2016wavenet, kalchbrenner2018efficient}. Specifically, we employ the Roformer network \cite{su2021roformer} for coarse-path processing, where we use a self-attention layer followed by a cross-attention layer to be conditional on $\mathbf{E}_\text{ST}$ with rotary positional embedding. On the other hand, we use a stack of 2-layer simple recurrent units (SRUs) \cite{lei2017simple} for fine-path processing. The feature-wise linear modulation (FiLM) \cite{perez2018film} is applied to the output of SRUs to assist the denoising with the angle embedding $\mathbf{E}_{\boldsymbol{\delta}}$ and the pooled $\mathbf{E}_\text{ST}$. Each of these processing stages is detailed below.

\paragraph{Coarse-Path Processing} 
In a dual-path block, we first process the coarse path corresponding to the vertical axis shown in Figure~\ref{fig:segmentation}, in which the columns are processed in parallel:
\begin{align}
\mathbb{H}^{(i)}_\text{c-out} := \text{RepeatSegments}\left(\left[\text{Roformer}\left(\text{MergeSegments}\left(\mathbb{H}^{(i)}\right)[:, k, :]\right), k=0, \ldots, K^{(i)}_\text{MS}-1\right]\right),
\end{align}
where the coarse-path output $\mathbb{H}^{(i)}_\text{c-out}\in\mathbb{R}^{S\times K\times D_\text{hid}}$ has the same shape as $\mathbb{H}^{(i)}$, and $\text{MergeSegments}(\cdot)$ and $\text{RepeatSegments}(\cdot)$ are the operations that, respectively, compress and expand the segments horizontally to aggregate the information within a segment for a coarser scale of inter-segment processing. Note that, without taking the merging and repeating operations, the vertical axis is simply a sequence formed by skipping $K/2$ elements in $\mathbf{H}$, which does not really capture the desired coarse information. The merging is done by averaging every pair of $2^{\min\{i, N-i+1\}}$ columns with zero paddings and a half stride such that $K^{(i)}_\text{MS}=\left\lceil \frac{K}{2^{\min\{i, N-i+1\}-1}}\right\rceil$. The upper part of Figure~\ref{fig:chunk-ops} illustrates the case of $i=2$. Similar to \cite{lam2021sandglasset}, our definition of $K^{(i)}_\text{MS}$ changes the width of the 3d tensor with the block index $i$ in a sandglass style, as we have the shortest segment at the middle block and the longest segment at the first and the last block. To match with the original length, a repeating operation following from the Roformer is performed, as shown in the lower part of Figure~\ref{fig:chunk-ops}.

\paragraph{Fine-Path Processing} We then obtain the fine-path input:
$
    \mathbb{H}^{(i)}_\text{f-in}:=\text{RMSNorm}\left(\mathbb{H}^{(i)}+\mathbb{H}^{(i)}_\text{c-out}\right),
$
which is fed to a two-layer SRU by parallelly processing the rows illustrated in Figure~\ref{fig:segmentation}:
\begin{align}
\mathbb{H}^{(i)}_\text{f-out} := \left[\text{FiLM}\left(\text{SRU}\left(\mathbb{H}^{(i)}_\text{f-in}[s, :, :]\right), 
\mathbf{E}_{\boldsymbol{\delta}}\left[\frac{sL}{S}, :\right]+\frac{1}{T_\text{ST}}\sum_{t=0}^{T_\text{ST}-1}\mathbf{E}_\text{ST}[t, :]\right), s=0, \ldots, S-1\right],
\end{align}
where $\text{FiLM}(\mathbf{x}, \mathbf{m}):=\text{MLP}_3\left(\left(\mathbf{x}\otimes\text{MLP}_1(\mathbf{m})\right)+\text{MLP}_2(\mathbf{m})\right)$ for an arbitrary input $\mathbf{x}$ and modulation condition $\mathbf{m}$, and $\otimes$ is the operations of broadcast multiplication. Followed from this, we have the input for the next dual-path block:
$
    \mathbb{H}^{(i+1)}:=\text{RMSNorm}\left(\mathbb{H}^{(i)}_\text{f-in}+\mathbb{H}^{(i)}_\text{f-out}\right).
$
After recursively processing through $N$ dual-path blocks, the 3-D tensor is transformed back to a 2-D matrix using an overlap-and-add method \cite{luo2020dual}. Finally, the predicted velocity is obtained as follows:
\begin{align}
\hat{\mathbf{v}}_\theta(\mathbf{z}_\text{noisy};\mathbf{c}):=\text{RMSNorm}\left(\text{OverlapAdd}\left(\mathbb{H}^{(N+1)}\right)\right)\mathbf{W}_\text{out},
\end{align}
where $\mathbf{W}_\text{out}\in\mathbb{R}^{D_\text{hid}\times D}$ is learnable. We present more details of our implementation in Appendix B.

\subsection{Audio VAE-GANs for Latent Representation Learning}
\label{sec:kl-vae}

To avoid learning arbitrarily high-variance latent representations, \citet{rombach2022high} examined a KL-regularized image autoencoder for latent diffusion models (LDMs) and demonstrated extraordinary stability in generating high-quality image \cite{stablediffusionv21}, igniting a series of follow-up works \cite{zhang2023adding}. Such an autoencoder imposes a KL penalty on the encoder outputs in a way similar to VAEs \cite{kingma2013auto, rezende2014stochastic}, but, different from the classical VAEs, it is adversarially trained as in the generative adversarial networks (GANs) \cite{creswell2018generative}. In this paper, this class of autoencoders is referred to as the {VAE-GAN}. Although VAE-GANs are promisingly applied to image generation, there is still a lack of comparable successful methods for the autoencoding of audio waveforms. In this work, we propose a similarly trained audio VAE-GAN, which empirically showed remarkable stability when applied to our DPD model in comparison to other commonly used VQ-VAE used in \cite{yang2023diffsound, liu2023audioldm, kreuk2022audiogen}.

Specifically, the audio VAE-GAN is trained to reconstruct 24kHz audio with a striding factor of 96, resulting in a 250Hz latent sequence. The architecture of the decoder is the same as that in HiFi-GAN \cite{kong2020hifi}. For the encoder, we basically replace the up-sampling modules in the decoder with convolution-based down-sampling modules while other modules stay the same. For adversarial training, we use the multi-period discriminators in \cite{kong2020hifi} and the multi-resolution spectrogram discriminators in \cite{jang2021univnet}. The training details are further discussed in Appendix B.
% Similar to the implementation in \cite{stablediffusionv21}, we use a KL loss to regularize the encoder outputs, denoted by $\bar{\mathbf{z}}$: $\text{KL}\left(\mathcal{N}(\bar{\mathbf{z}}; \boldsymbol{\mu}, \boldsymbol{\sigma}\mathbf{I})||\mathcal{N}(\mathbf{0}, \mathbf{I})\right)$.
To match the normal range of targets for diffusion models \cite{ho2020denoising, rombach2022high}, we map the encoder outputs to $[-1, 1]$ by ${\mathbf{z}}_{(i,j)}:=\min\left\{\max\left\{\bar{\mathbf{z}}_{(i,j)}/3, -1\right\}, 1\right\} \forall i, j$, where the subscript $(i,j)$ denotes the value on the $i$-th row and $j$-th column, and the choice of $3$ in practice would sieve extreme values occupying $<0.1\%$. 

\subsection{Music Inpainting, Music Continuation and Music Prompting with MeLoDy}
We show that the proposed MeLoDy supports interpolation (i.e., audio inpainting) and extrapolation (i.e., audio continuation) with tricks of manipulating random noises. Noticeably, diffusion models have been successfully used for effective audio inpainting \cite{liu2023audioldm, huang2023make}. Yet, audio continuation has been an obstacle for diffusion models due to their non-autoregressive nature.
Besides audio continuation, based on MuLan, MeLoDy also supports music prompts to generate music of a similar style, as shown in Figure~\ref{fig:overall}. Examples of music inpainting, music continuation, and music prompting are shown on our demo page. We present the algorithms of these functionalities in Appendix C.

% Distinctive from MusicLM, in the DPD pipeline, we also take advantage of the reference audio $\mathbf{z}_\text{ref}$ to directly sample a noisy latent $\mathbf{z}_{\delta_{T'}}\sim q(\mathbf{z}_{\delta_{T'}}|\mathbf{z}=\mathbf{z}_\text{ref})$, and use it as the starting point of sampling in place of random noise. While the MuLan mainly extracts the timbre of the audio, 

% \begin{align}
%     \mathbf{z}_{{\delta_t}-\omega_t}=\cos(\omega_t)\mathbf{z}_{\delta_t}-\sin(\omega_t)\hat{\mathbf{v}}_\theta(\mathbf{z}_{\delta_t};\left\{\mathbf{u}_{\lceil T_\text{ST}/M\rceil}, \ldots, \mathbf{u}_{T_\text{ST}+\lceil T_\text{ST}/M\rceil}, \boldsymbol\delta_\text{new}\right\}).
% \end{align}

\section{Experiments}
\label{sec:exps}

\begin{table}
  \caption{The speed and the quality of our proposed MeLoDy on a CPU (Intel Xeon Platinum 8260 CPU @ 2.40GHz) or a GPU (NVIDIA Tesla V100) using different numbers of sampling steps.}
  \label{tab:speed-vs-quality}
  \centering
  \begin{tabular}{ccccc}
    \toprule
    % \multicolumn{2}{c}{Part}                   \\
    % \cmidrule(r){1-2}
    \textbf{Steps} ($T$) & \textbf{Speed on CPU} ($\uparrow$) & \textbf{Speed on GPU} ($\uparrow$) & \textbf{FAD}  ($\downarrow$) & \textbf{MCC} ($\uparrow$) \\
    \midrule
    (MusicCaps) & - & - & - & 0.43 \\
    \midrule
    5 & \textbf{1472Hz (0.06$\times$)} & \textbf{181.1kHz (7.5$\times$)} & 7.23 & 0.49\\
    10 & 893Hz (0.04$\times$) & 104.8kHz (4.4$\times$) & 5.93 & 0.52\\
    20 & 498Hz (0.02$\times$) & 56.9kHz (2.4$\times$) & \textbf{5.41} & \textbf{0.53} \\
    \bottomrule
  \end{tabular}
  \vspace{-0.68em}
\end{table}

\subsection{Experimental Setup}

\paragraph{Data Preparation} As shown in Table~\ref{tab:previous-works}, MeLoDy was trained on 257k hours of music data (6.4M 24kHz audios), which were filtered with \cite{kong2020panns} to focus on non-vocal music. Additionally, inspired by the text augmentation in \cite{huang2023noise2music}, we enriched the tag-based texts to generate music captions by asking ChatGPT \cite{chatgpt}. This music description pool is used for the training of our 195.3M MuLan, where we randomly paired each audio with either the generated caption or its respective tags. In this way, we robustly improve the model's capability of handling free-form text.

\paragraph{Semantic LM}
For semantic modeling, we trained a 429.5M LLaMA \cite{touvron2023llama} with 24 layers, 8 heads, and 2048 hidden dimensions, which has a comparable number of parameters to that of the MusicLM \cite{agostinelli2023musiclm}. For conditioning, we set up the MuLan RVQ using 12 1024-sized codebooks, resulting in 12 prefixing tokens. The training targets were 10s semantic tokens, which are obtained from discretizing the 25Hz embeddings from a 199.5M Wav2Vec2-Conformer with 1024-center k-means.

\paragraph{Dual-Path Diffusion} For the DPD model, we set the hidden dimension to $D_\text{hid}=768$, and block number to $N=8$, resulting in 296.6M parameters. For the input chunking strategy, we divide the 10s training inputs in a fixed length of $L=2500$ into $M=4$ parts. For segmentation, we used a segment size of $K=64$ (i.e., each segment is 256ms long), leading to $S=80$ segments. In addition, we applied the classifier-free guidance \cite{ho2022classifier} to DPD for improving the correspondence between samples and conditions. During training, the cross-attention to semantic tokens is randomly replaced by self-attention with a probability of $0.1$. For sampling, the predicted velocity is linearly combined as . For all of our generations, a scale of 2.5 was used for classifier-free guidance.

\paragraph{Audio VAE-GAN} For audio VAE-GAN, we used a hop size of 96, resulting in 250Hz latent sequences for encoding 24kHz music audio. The latent dimension $D=16$, thus we have a total compression rate of 6$\times$. The hidden channels used in the encoder were 256, whereas that used in the decoder were 768. The audio VAE-GAN in total contains 100.1M parameters.

\subsection{Performance Analysis}
\paragraph{Objective Metrics} We use the VGGish-based \cite{hershey2017cnn} Fre\'{c}het audio distance (FAD) \cite{kilgour2019frechet} between the generated audios and the reference audios from MusicCaps \cite{agostinelli2023musiclm} as a rough measure of generation fidelity.\footnote{Note that MeLoDy was mainly trained with non-vocal music data, its sample distribution could not fit the reference one as well as in \cite{agostinelli2023musiclm, huang2023noise2music}, since about 76\% audios in MusicCaps contain either vocals or speech.} To measure text correlation, we use the MuLan cycle consistency (MCC) \cite{agostinelli2023musiclm}, which calculates the cosine similarity between text and audio embeddings using a pre-trained MuLan.\footnote{Since our MuLan model was trained with a different dataset, our MCC results cannot be compared to \cite{agostinelli2023musiclm, huang2023noise2music}.}

\paragraph{Inference Speed}
We first evaluate the sampling efficiency of our proposed MeLoDy. As DPD permits using different numbers of sampling steps depending on our needs, we report its generation speed in Table~\ref{tab:speed-vs-quality}. Surprisingly, MeLoDy  steadily achieved a higher MCC score than that of the reference set, even taking only 5 sampling steps. This means that (i) the MuLan model determined that our generated samples were more correlated to MusicCaps captions than reference audios, and (ii) the proposed DPD is capable of consistently completing the MuLan cycle at significantly lower costs than the nested LMs in \cite{agostinelli2023musiclm}.

\paragraph{Comparisons with SOTA models} 
We evaluate the performance of MeLoDy by comparing it to MusicLM \cite{agostinelli2023musiclm} and Noise2Music \cite{huang2023noise2music}, which both were trained large-scale music datasets and demonstrated SOTA results for a wide range of text prompts. To conduct fair comparisons, we used the same text prompts in their demos (70 samples from MusicLM; 41 samples from Noise2Music),\footnote{All samples for evaluation are available at {https://Efficient-MeLoDy.github.io/}. Note that our samples were not cherry-picked, whereas the samples we compared were cherry-picked \cite{huang2023noise2music}, constituting very strong baselines.} and asked seven music producers to select the best out of a pair of samples or voting for a tie (both win) in terms of musicality, audio quality, and text correlation. In total, we conducted 777 comparisons and collected 1,554 ratings. We detail the evaluation protocol in Appendix F. Table~\ref{tab:performance} shows the comparison results, where each category of ratings is separated into two columns, representing the comparison against MusicLM (MLM) or Noise2Music (N2M), respectively. Finally, MeLoDy consistently achieved comparable performances (all winning proportions fall into [0.4, 0.6]) in musicality and text correlation to MusicLM and Noise2Music. Regarding audio quality, MeLoDy outperformed MusicLM ($p<0.05$) and Noise2Music ($p<0.01$), where the $p$-values were calculated using the Wilcoxon signed-rank test. We note that, to sample 10s and 30s music, MeLoDy only takes 4.32\% and 0.41\% NFEs of MusicLM, and 10.4\% and 29.6\% NFEs of Noise2Music, respectively.
% \footnote{}
% Concurrent with this work, we noticed that the MusicLM API was released at {https://aitestkitchen.withgoogle.com/experiments/music-lm}, where the samples appeared to be weaker than the demo ones. Yet, we decide to keep the demo samples as the comparing targets for not to weaken our baselines.

\paragraph{Diversity Analysis} Diffusion models are distinguished for its high diversity \cite{xiao2021tackling}. We conduct an additional experiment to study  the diversity and validity of MeLoDy's generation given the same text prompt of open description, e.g., feelings or scenarios. The sampled results were shown on our demo page, in which we obtained samples with diverse combinations of instruments and textures.

\paragraph{Ablation Studies}
We also study the ablation on two aspects of the proposed method. In Appendix D, we compared the uniform angle schedule in \cite{schneider2023mo} and the linear one proposed in DPD using the MCC metric and case-by-case qualitative analysis. It turns out that our proposed schedule tends to induce fewer acoustic issues when taking a small number of sampling steps. In Appendix E, we showed that the proposed dual-path architecture outperformed other architectures \cite{schneider2023mo, riffusion} used for LDMs in terms of the signal-to-noise ratio (SNR) improvements using a subset of the training data.

\begin{table}
  \caption{The comparison of MeLoDy with the SOTA text-to-music generation models. \textbf{NFE} is the number of function evaluations \cite{salimans2022progressive} for generating $T$-second audio.\tablefootnote{We use $+$ to separate the counts for the iterative modules, i.e., LM or DPM. Suppose the cost of each module is comparable, then the time steps taken by LM and the diffusion steps taken by DPM can be fairly compared.} \textbf{Musicality}, \textbf{Quality}, and \textbf{Text Corr.} are the winning proportions in terms of musicality, quality, and text correlation, respectively.}
  \label{tab:performance}
  \centering
  \begin{tabular}{lccccccc}
    \toprule
    % \multicolumn{2}{c}{Part}                   \\
    % \cmidrule(r){1-2}
    \multirow{2}{*}{\textbf{Model}} & \multirow{2}{*}{\textbf{NFE} ($\downarrow$)} &     \multicolumn{2}{c}{\textbf{Musicality} ($\uparrow$)} &     \multicolumn{2}{c}{\textbf{Quality} ($\uparrow$)} &     \multicolumn{2}{c}{\textbf{Text Corr.} ($\uparrow$)}  \\
    \cmidrule(r){3-8}
    & & MLM & N2M & MLM & N2M & MLM & N2M
    \\
    
    \midrule
    MusicLM \cite{agostinelli2023musiclm} & $(25+200+400)T$ & \textbf{0.541} & - & {0.465} & - & \textbf{0.548} & -\\
    Noise2Music \cite{huang2023noise2music} & $1000+800+800$  & - & \textbf{0.555} & - & 0.436 & - & \textbf{0.572} \\
    \midrule
    \textbf{MeLoDy} (20 steps) & $25T+20$ & 0.459 & 0.445 & \textbf{0.535} & \textbf{0.564}  & {0.452} & 0.428 \\
    \bottomrule
  \end{tabular}
  \vspace{-0.68em}
\end{table}

\section{Discussion}

\paragraph{Limitation}
We acknowledge the limitations of our proposed MeLoDy. To prevent from having any disruption caused by unnaturally sound vocals, our training data was prepared to mostly contain non-vocal music only, which may limit the range of effective prompts for MeLoDy. Besides, the training corpus we used was unbalanced and slightly biased towards pop and classical music. Lastly, as we trained the LM and DPD on 10s segments, the dynamics of a long generation may be limited.

\paragraph{Broader Impact}
We believe our work has a huge potential to grow into a music creation tool for music producers, content creators, or even normal users to seamlessly express their creative pursuits with a low entry barrier. MeLoDy also facilitates an interactive creation process, as in Midjourney \cite{holz2022midjourney}, to take human feedback into account. For a more precise tune of MeLoDy on a musical style, the LoRA technique \cite{hu2021lora} can be potentially applied to MeLoDy, as in Stable Diffusion \cite{stablediffusionv21}.

\medskip

\bibliographystyle{unsrtnat}
\bibliography{neurips_2023}

\import{./}{appendix.tex}

\end{document}

%% file: appendix.tex
\newpage
\appendix
\section{Mathematical Background for Dual-Path Diffusion}
\subsection{Forward Diffusion Process}
In dual-path diffusion (DPD), we consider a Gaussian diffusion process \cite{kingma2021variational} that continuously diffuses our generation target $\mathbf{z}$ into increasingly noisy versions of $\mathbf{z}$, denoted as $\mathbf{z}_t$ with $t \in [0, 1]$ running from $t = 0$ (least noisy) to $t = 1$ (most noisy). This forward diffusion process is formally defined as
\begin{align}
    q(\mathbf{z}_t|\mathbf{z})=\mathcal{N}(\mathbf{z}_t; \alpha_t \mathbf{z}, \sigma_t^2\mathbf{I}),
    \label{eq:prior}
\end{align}
where two strictly positive scalar-valued, continuously differentiable functions $\alpha_t, \sigma_t$ define the noise schedule \cite{ho2020denoising} of this forward diffusion process. Building upon the nice properties of Gaussian distributions, we can express $q(\mathbf{z}_t|\mathbf{z}_s)$, for any $0 \leq s < t \leq 1$, as another Gaussian distribution:
\begin{align}
    q(\mathbf{z}_t|\mathbf{z}_s)=\mathcal{N}\left(\mathbf{z}_t; \frac{\alpha_t}{\alpha_s} \mathbf{z}_s, \left(\sigma_t^2-\frac{\alpha_t}{\alpha_s}\sigma_s^2\right)\mathbf{I}\right).
\label{eq:likelihood}
\end{align}
Regarding the choice of noise scheduling functions, we consider the typical setting used in \cite{jascha2015, ho2020denoising}: $\alpha_t=\sqrt{1-\sigma_t^2}$, which gives rise to a \textit{variance-preserving} diffusion process \cite{kingma2021variational}. Specifically, we employ the trigonometric functions in \cite{salimans2022progressive}, defined as follows:
\begin{align}
    &\alpha_t:=\cos(\pi t/2)\,\,\,\,\,\sigma_t:=\sin(\pi t/2)\,\,\,\,\,\forall t\in[0, 1]\\
    \Leftrightarrow \,\,\,&\alpha_\delta:=\cos(\delta)\,\,\,\,\,\,\,\,\,\,\sigma_\delta:=\sin(\delta)\,\,\,\,\,\,\,\,\,\,\forall \delta\in[0, \pi/2].
\end{align}
With this re-parameterization, the diffusion process can now be defined in terms of angle $\delta\in [0, \pi/2]$:
\begin{align}
\mathbf{z}_{\delta}=\cos(\delta)\mathbf{z}+\sin(\delta)\boldsymbol\epsilon,\,\,\,\,\,\boldsymbol\epsilon\sim\mathcal{N}(\mathbf{0}, \mathbf{I}),
\end{align}
where $\mathbf{z}_\delta$ gets noisier as $\delta$ increases from $0$ to $\pi/2$.
\subsection{Prediction of Diffusion Velocity}
The diffusion velocity of $\mathbf{z}_{\delta}$ at $\delta$ \cite{salimans2022progressive} is defined as:
\begin{align}
    \mathbf{v}_\delta:=\frac{d \mathbf{z}_{\delta}}{d \delta}=\frac{d \cos({\delta})}{d \delta}\mathbf{z}+\frac{d \sin({\delta})}{d \delta}\boldsymbol\epsilon=\cos({\delta})\boldsymbol\epsilon-\sin({\delta})\mathbf{z}.
\end{align}
Based on $\mathbf{v}_\delta$, we can compute $\mathbf{z}$ and $\boldsymbol\epsilon$ from a noisy latent $\mathbf{z}_\delta$:
\begin{align}
    &\mathbf{z}=\cos{(\delta)}\mathbf{z}_\delta-\sin({\delta})\mathbf{v}_\delta=\alpha_\delta \mathbf{z}_\delta-\sigma_\delta\mathbf{v}_\delta;\label{eq:z-from-v}
    \\
    &\boldsymbol\epsilon=\sin{(\delta)}\mathbf{z}_\delta+\cos({\delta})\mathbf{v}_\delta=\sigma_\delta\mathbf{z}_\delta+\alpha_\delta\mathbf{v}_\delta,
    \label{eq:e-from-v}
\end{align}
which suggests $\mathbf{v}_\delta$ a feasible target for network prediction $\hat{\mathbf{v}}_\theta(\mathbf{z}_\delta;\mathbf{c})$ given a collection of conditions $\mathbf{c}$, as an alternative to the $\mathbf{z}$ prediction ($\hat{\mathbf{z}}_\theta(\mathbf{z}_\delta;\mathbf{c})$), e.g., in \cite{kingma2021variational}, and the $\boldsymbol\epsilon$ prediction ($\hat{\boldsymbol\epsilon}_\theta(\mathbf{z}_\delta;\mathbf{c})$), e.g., in \cite{ho2020denoising, kong2020diffwave, chen2020wavegrad}. As reported by \citet{salimans2022progressive} and \citet{schneider2023mo}, training the neural network $\theta$ with a mean squared error (MSE) loss as in the pioneering work \cite{ho2020denoising} remains effective:
\begin{align}
    \mathcal{L}:=\mathbb{E}_{\mathbf{z}\sim p_\text{data}(\mathbf{z}), {\boldsymbol\epsilon}\sim\mathcal{N}(\mathbf{0}, \mathbf{I}), \delta\sim\text{Uniform}[0, 1]}\left[ \left\lVert\cos({\delta})\boldsymbol\epsilon-\sin({\delta})\mathbf{z}-\hat{\mathbf{v}}_\theta(\cos(\delta)\mathbf{z}+\sin(\delta)\boldsymbol\epsilon;\mathbf{c})\right\rVert^2_2\right],
\end{align}
which forms the basis of DPD's training loss, i.e., the simplest case of considering only a single chunk per input ($M=1$) in Eq. (\ref{eq:diff-loss}). We can easily extend this to a multi-chunk version by sampling $M$ different angles $\delta_1, \ldots, \delta_M \sim\text{Uniform}[0, 1]$, where the $m$-th sampled angle is applied to the corresponding chunk of the latent, i.e., $\mathbf{z}[(m-1)L/M:mL/M]$.

\subsection{Generative Diffusion Process}
Generation is done by inverting the forward process from a noise vector randomly drawn from $\mathcal{N}(\mathbf{0}, \mathbf{I})$.
One efficient way to accomplish this is to take advantage of DDIM \cite{song2020denoising}, which enables running backward from angle $\delta$ to angle $\delta-\omega$, for any step size $0< \omega<\delta$:
\begin{align}
p_\theta(\mathbf{z}_{\delta-\omega}|\mathbf{z}_{\delta}):=q\left(\mathbf{z}_{\delta-\omega}\left|\mathbf{z}=\frac{\mathbf{z}_{\delta}-\sigma_\delta \hat{\boldsymbol\epsilon}_\theta(\mathbf{z}_\delta;\mathbf{c})}{\alpha_\delta}\right.\right)=\alpha_{\delta-\omega}\left(\frac{\mathbf{z}_{\delta}-\sigma_\delta \hat{\boldsymbol\epsilon}_\theta(\mathbf{z}_\delta;\mathbf{c})}{\alpha_\delta}\right)+\sigma_{\delta-\omega}\boldsymbol{\epsilon},
\end{align}
where $\boldsymbol{\epsilon}\sim \mathcal{N}(\mathbf{0}, \mathbf{I})$. \citet{song2020denoising} considered $\boldsymbol{\epsilon}\equiv\hat{\boldsymbol\epsilon}_\theta(\mathbf{z}_\delta;\mathbf{c})$, leading to a deterministic update rule:
\begin{align}
\mathbf{z}_{\delta-\omega}&=\frac{\alpha_{\delta-\omega}}{\alpha_\delta}\mathbf{z}_{\delta}+\left(\sigma_{\delta-\omega}-\frac{\alpha_{\delta-\omega}\sigma_\delta}{\alpha_\delta}\right)\hat{\boldsymbol\epsilon}_\theta(\mathbf{z}_\delta;\mathbf{c}).
\end{align}
Building upon the diffusion velocity, \citet{salimans2022progressive} re-parameterized DDIM as
\begin{align}
    p_\theta(\mathbf{z}_{\delta-\omega}|\mathbf{z}_{\delta}):=&q\left(\mathbf{z}_{\delta-\omega}\left|\mathbf{z}=\alpha_\delta \mathbf{z}_\delta-\sigma_\delta\hat{\mathbf{v}}_\theta(\mathbf{z}_\delta;\mathbf{c})\right.\right)\\
    =&\alpha_{\delta-\omega}\left(\alpha_\delta \mathbf{z}_\delta-\sigma_\delta\hat{\mathbf{v}}_\theta(\mathbf{z}_\delta;\mathbf{c})\right)+\sigma_{\delta-\omega}\boldsymbol{\epsilon},
\end{align}
where $\boldsymbol{\epsilon}\sim \mathcal{N}(\mathbf{0}, \mathbf{I})$. Here, we can similarly consider a parameterized noise vector $\boldsymbol{\epsilon}\equiv\sigma_{\delta}\mathbf{z}_\delta+\alpha_{\delta}\hat{\mathbf{v}}_\theta(\mathbf{z}_\delta;\mathbf{c})$ based on Eq. (\ref{eq:e-from-v}), yielding a simplified deterministic update rule:
\begin{align}
    \mathbf{z}_{\delta-\omega}=&\alpha_{\delta-\omega}\left(\alpha_\delta \mathbf{z}_\delta-\sigma_\delta\hat{\mathbf{v}}_\theta(\mathbf{z}_\delta;\mathbf{c})\right)+\sigma_{\delta-\omega}\left(\sigma_{\delta}\mathbf{z}_\delta+\alpha_{\delta}\hat{\mathbf{v}}_\theta(\mathbf{z}_\delta;\mathbf{c})\right)\\
    =&\left(\alpha_{\delta-\omega}\alpha_\delta-\sigma_{\delta-\omega}\sigma_\delta\right)\mathbf{z}_\delta+\left(\sigma_{\delta-\omega}\alpha_\delta-\alpha_{\delta-\omega}\sigma_\delta\right)\hat{\mathbf{v}}_\theta(\mathbf{z}_\delta;\mathbf{c})\\
    =&\cos({\omega})\mathbf{z}_\delta-\sin({\omega})\hat{\mathbf{v}}_\theta(\mathbf{z}_\delta;\mathbf{c})
\end{align}
where the last equation is obtained by applying the trigonometric identities:
\begin{align}
\alpha_{\delta-\omega}\alpha_\delta-\sigma_{\delta-\omega}\sigma_\delta&=\cos(\delta-\omega)\cos(\delta)-\sin(\delta-\omega)\sin(\delta)=\cos({\omega});\\
\sigma_{\delta-\omega}\alpha_\delta-\alpha_{\delta-\omega}\sigma_\delta&=\sin(\delta-\omega)\cos(\delta)-\cos(\delta-\omega)\sin(\delta)=\sin({\omega}).
\end{align}
Building upon this angular update rule and having specified the angle step sizes $\omega_1, \ldots, \omega_T$ with $\sum_{t=1}^{T}\omega_t=\pi/2$, we can generate samples from $\mathbf{z}_{\pi/2}\sim\mathcal{N}(\mathbf{0}, \mathbf{I})$ after $T$ steps of sampling:
\begin{align}
    \mathbf{z}_{\delta_{t}-\omega_t}=\cos({\omega_t})\mathbf{z}_{\delta_t}-\sin({\omega_t})\hat{\mathbf{v}}_\theta(\mathbf{z}_{\delta_t};\mathbf{c}),\,\,\,\,\,\delta_{t}=\begin{cases}
    \frac{\pi}{2}-\sum_{i=t+1}^T\omega_i, & 1 \leq t < T;\\
    \frac{\pi}{2}, & t = T,
\end{cases}
\end{align}
running from $t=T$ to $t=1$.

\section{Training and Implementation Details}
\subsection{Audio VAE-GAN}
\begin{figure}[h]
     \centering
     \includegraphics[width=0.6\textwidth]{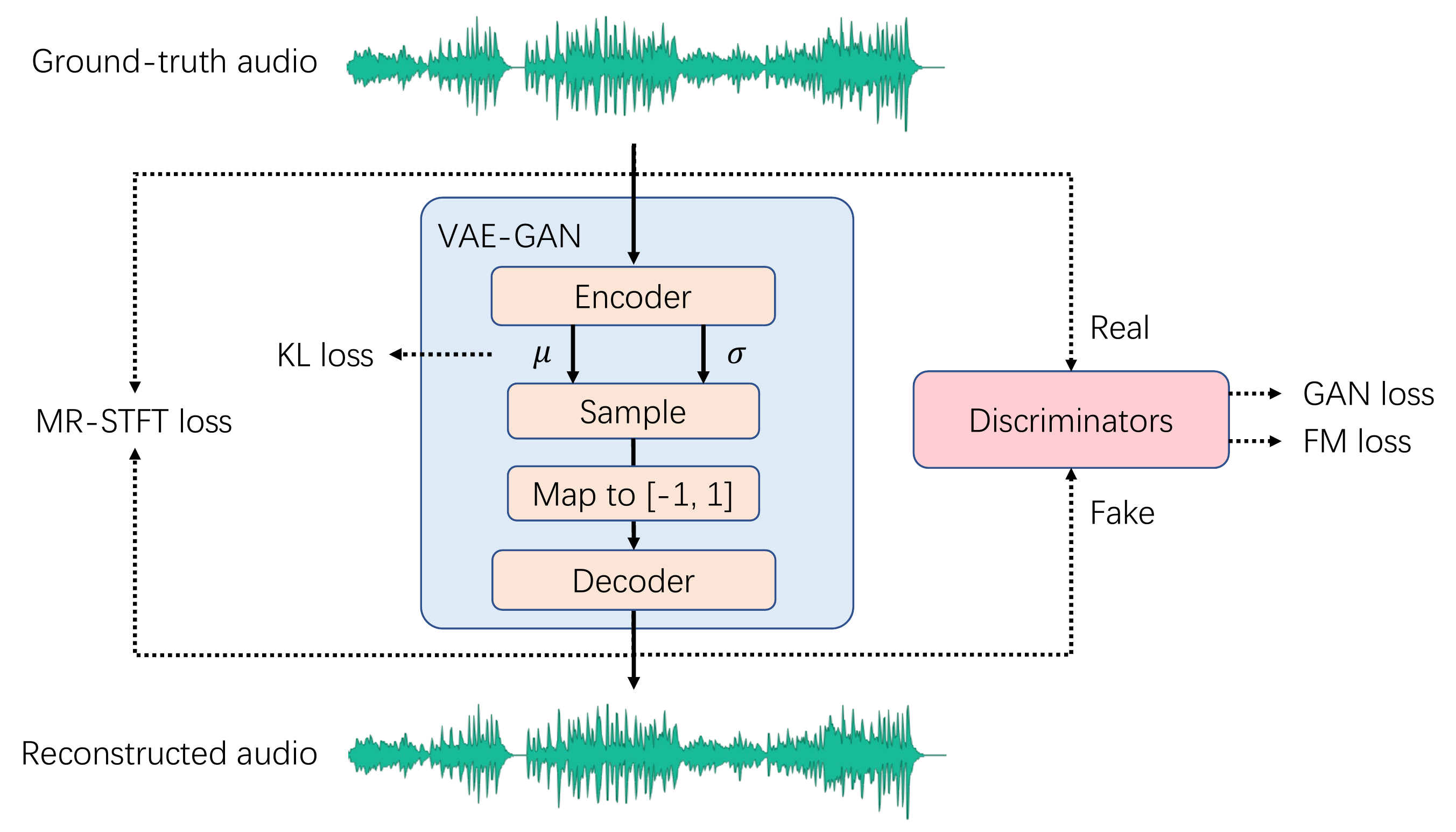}
     \caption{The audio VAE-GAN trained for dual-path diffusion models}
     \label{fig:vae}
\end{figure}
As shown in Figure \ref{fig:vae}, we train a VAE-GAN to extract 250Hz 16-dimensional latent $\mathbf{z}\in\mathbb{R}^{L\times 16}$ from a 24kHz audio $\mathbf{x}\in\mathbb{R}^{T_\text{wav}}$ with $L=\lceil T_\text{wav} / 96 \rceil$. The audio VAE-GAN mainly comprises three trainable modules: (i) a variational Gaussian encoder $\mathcal{E}_\phi(\mathbf{x})\equiv\mathcal{N}(\mu_\phi(\mathbf{x}), \sigma_\phi(\mathbf{x})\mathbf{I})$, (ii) a decoder $\mathcal{D}_\phi(\mathbf{z})$, and (iii) a set of $n$ discriminators $\{D^{(i)}\}_{i=1}^{n}$.

Regarding network architecture, we use the ResNet-style convolutional neural networks (CNNs) in HiFi-GAN \cite{kong2020hifi} as the backbone.\footnote{Our implementation is similar to that in https://github.com/jik876/hifi-gan.} For the encoder, we replace the up-sampling blocks in HiFi-GAN with convolution-based down-sampling blocks, with down-sampling rates of $[2, 3, 4, 4]$, output dimensions of $[32, 64, 128, 256]$ and kernel sizes of $[5, 7, 9, 9]$ in four down-sampling blocks, giving 40M parameters. The final layer of the encoder maps the 256-dimensional output to two 16-dimensional latent sequences, respectively for the mean and variance of diagonal Gaussian sampling.\footnote{The Gaussian sampling is referred to LDMs' implementation at https://github.com/CompVis/latent-diffusion/blob/main/ldm/modules/distributions/distributions.py} As shown in Figure \ref{fig:vae}, to match the normal range of targets for diffusion models \cite{ho2020denoising, rombach2022high}, we map the sampled outputs to $[-1, 1]$ by ${\mathbf{z}}_{(i,j)}:=\min\left\{\max\left\{\bar{\mathbf{z}}_{(i,j)}/3, -1\right\}, 1\right\} \forall i, j$, where the subscript $(i,j)$ denotes the value on the $i$-th row and $j$-th column, and the choice of $3$ in practice would sieve extreme values occupying $<0.1\%$. For the architecture setting of the decoder, it inherits the same architecture of HiFi-GAN, and uses up-sampling rates of $[4, 4, 3, 2]$, kernel sizes of $[9, 9, 5, 7]$ and larger number of output channels ($[768, 384, 192, 96]$) for four up-sampling blocks, taking 60.1M parameters.

For adversarial training, we use the multi-period discriminators in \cite{kong2020hifi} and the multi-resolution spectrogram discriminators in \cite{jang2021univnet}. The training scheme is similar to that in \cite{kong2020hifi}. The training loss for the encoder and the decoder comprises four components:
\begin{align}
    \mathcal{L}_\text{vae-gan}(\phi):=&\mathbb{E}_{\mathbf{x}\sim p_\text{data}(\textbf{x})}\left[\mathbb{E}_{\mathbf{z}\sim\mathcal{E}_\phi(\textbf{x})}\left[\lambda_\text{mr-stft}\mathcal{L}_\text{mr-stft}+\lambda_\text{fm}\mathcal{L}_\text{fm}+\lambda_\text{gan}\mathcal{L}_\text{gan}+\lambda_\text{kl}\mathcal{L}_\text{kl}\right]\right]\\
    \mathcal{L}_\text{mr-stft}:=&\sum_{r=1}^{R}\left\lVert\text{STFT}_r(\textbf{x})-\text{STFT}_r\left(\mathcal{D}_\phi\left(\mathbf{z}\right)\right)\right\rVert_1\\
    \mathcal{L}_\text{fm}:=&\sum_{i=1}^{n} \frac{1}{|D^{(i)}|}\sum_{l=1}^{|D^{(i)}|} \left\lVert D^{(i)}_l(\mathbf{x}) - D^{(i)}_l(\mathcal{D}_\phi\left(\mathbf{z}\right))\right\rVert_1\\
    \mathcal{L}_\text{gan}:=&\sum_{i=1}^n\left(D^{(i)}\left(\mathcal{D}_\phi\left(\mathbf{z}\right)\right)-1\right)^2\\
    \mathcal{L}_\text{kl}:=&\text{KL}\left(\mathcal{E}_\phi(\mathbf{x})||\mathcal{N}(\mathbf{0}, \mathbf{I})\right),
\end{align}
where $\text{STFT}_r$ computes the magnitudes after the $r$-th short-time Fourier transform (STFT) out of $R=7$ STFTs (the number of FFTs $=[8192, 4096, 2048, 512, 128, 64, 32]$; the window sizes $=[4096, 2048, 1024, 256, 64, 32, 16]$; the hop sizes $=[2048, 1024, 512, 128, 32, 16, 8]$), $|D^{(i)}|$ denotes the number of hidden layers used for feature matching in discriminator $D^{(i)}$, $D^{(i)}_l$ denotes the outputs of the $l$-th hidden layers in discriminator $D^{(i)}$, and $\lambda_\text{mr-stft}$, $\lambda_\text{fm}$, $\lambda_\text{gan}$, $\lambda_\text{kl}$ are the weights, respectively, for the multi-resolution STFT loss $\mathcal{L}_\text{mr-stft}$, the feature matching loss $\mathcal{L}_\text{fm}$, the GAN's generator loss $\mathcal{L}_\text{gan}$, and the Kullback–Leibler divergence based regularization loss $\mathcal{L}_\text{kl}$. To balance the scale of different losses, we set $\lambda_\text{mr-stft}=50$, $\lambda_\text{fm}=20$, $\mathcal{L}_\text{gan}=1$, and $\lambda_\text{kl}=5\times 10^{-3}$ in our training. In practice, we find it critical to lower the scale of the KL loss for a better reconstruction, though the distribution of the latents can still be close to zero mean and unit variance. 

\subsection{Wav2Vec2-Conformer}
Our implementation of Wav2Vec2-Conformer was based on an open-source library.\footnote{https://huggingface.co/docs/transformers/model\_doc/wav2vec2-conformer} In particular, Wav2Vec2-Conformer follows the same architecture as Wav2Vec2 \cite{baevski2020wav2vec}, but replaces the Transformer structure with the Conformer \cite{gulati2020conformer}. This model with 199.5M parameters was trained in self-supervised learning (SSL) manner similar to \cite{baevski2020wav2vec} using our prepared 257k hours of music data. 

\subsection{MuLan}
Our reproduced MuLan \cite{huang2022mulan} is composed of a music encoder and a text encoder. For music encoding, we rely on a publicly accessible Audio Spectrogram Transformer (AST) model pre-trained on AudioSet,\footnote{https://huggingface.co/MIT/ast-finetuned-audioset-10-10-0.4593} which gives promising results on various audio classification benchmarks. For text encoding, we employ the BERT \cite{devlin2018bert} base model pre-trained on a large corpus of English data using a masked language modeling (MLM) objective.\footnote{https://huggingface.co/bert-base-uncased} These two pre-trained encoders, together having 195.3M parameters, were subsequently fine-tuned on the 257k hours of music data with a text augmentation technique similar to \cite{huang2023noise2music}. In particular, we enriched the tag-based texts to generate music captions by asking ChatGPT \cite{chatgpt}. At training time, we randomly paired each audio with either the generated caption or its respective tags. In practice, this could robustly improve the model's capability of handling free-form text.

\section{Algorithms for MeLoDy}

\import{./}{algo.tex}

MeLoDy supports music or text prompting for music generation, as illustrated in Figure \ref{fig:overall}. We concretely detail the sampling procedures in Algorithm \ref{alg:music-gen}, where the algorithm starts by generating the latent sequence of length $L$ and then recursively prolongs the latent sequence using Algorithm \ref{alg:music-continue} until it reaches the desired length.

We further explain how music continuation can be effectively done in DPD.
Recall that the inputs for training DPD are the concatenated chunks of noisy latents in different noise scales. To continue a given music audio, we can add a new chunk composed of random noises and drop the first chunk. This is feasible since the conditions (i.e., the semantic tokens and the angles) defined for DPD have an autoregressive nature. Based on the semantic LM, we can continue the generation of $\lceil T_\text{ST}/M\rceil$ semantic tokens for the new chunk. Besides, it is sensible to keep the chunks other than the new chunk to have zero angles: $\boldsymbol\delta_\text{new}:=[0]_{r=1}^{L-\lceil L/M\rceil}\oplus[\delta_t]_{r=1}^{\lceil L/M\rceil}$, as shown in Algorithm \ref{alg:music-continue}.

In addition, music inpainting can be done in a similar way. We replace the inpainting partition of the input audio with random noise and partially set the angle vector to zeros to mark the positions where the denoising operations are not needed. Yet, in this case, the semantic tokens can only be roughly estimated using the remaining part of the music audio. 

\begin{table}[ht]
  \caption{The objective measures for the ablation study on angle schedules.}
  \label{tab:ablation-angle}
  \centering
  \begin{tabular}{cccc}
    \toprule
    % \multicolumn{2}{c}{Part}                   \\
    % \cmidrule(r){1-2}
    \textbf{Angle schedule} ($\omega_t$) & \textbf{Steps} ($T$) & \textbf{FAD}  ($\downarrow$) & \textbf{MCC} ($\uparrow$)\\
    \midrule
    \multirow{2}{*}{Uniform \cite{schneider2023mo}: $\omega_t=\frac{\pi}{2T}$} & 10 & 8.52 & 0.45 \\
    & 20 & 6.31 & 0.49 \\
    % \cmidrule{2-4}
    \midrule
    \multirow{2}{*}{Ours proposed in Eq. (\ref{eq:angle-schedule}): $\omega_t=\frac{\pi}{6T}+\frac{2\pi t}{3T(T+1)}$} & 10 & \textbf{5.93} & \textbf{0.52}  \\
    & 20 & \textbf{5.41} & \textbf{0.53} \\
    \bottomrule
  \end{tabular}
\end{table}

\section{Ablation Study on Angle Schedules}

We conduct an ablation study on angle schedules to validate the effectiveness of our proposed angle schedule $\omega_1, \ldots, \omega_T$ in Eq. (\ref{eq:angle-schedule}) in comparison to the previous uniform angle schedule \cite{schneider2023mo} also used for angle-parameterized continuous-time diffusion models. In particular, the same pre-trained DPD model $\hat{\mathbf{v}}_\theta$ and was used to sample with two different angle schedules with 10 steps and 20 steps, respectively, conditional on the same semantic tokens generated for the text prompts in MusicCaps. Table \ref{tab:ablation-angle} shows their corresponding objective measures in terms of FAD and MCC. We observe a significant improvement, especially when taking a small number of sampling steps, by using the proposed sampling method. This is aligned with our expectations that taking larger steps at the beginning of the sampling followed by smaller steps could improve the quality of samples, similar to the findings in previous diffusion scheduling methods \cite{chen2020wavegrad,lam2022bddm}.

\begin{figure}[ht]
     \centering
     \begin{subfigure}[b]{0.49\textwidth}
         \centering
         \includegraphics[width=\textwidth]{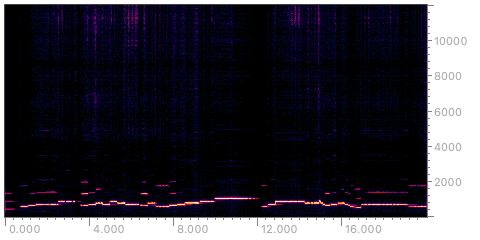}
         \caption{10-step sampling with uniform angle schedule for text prompt: ``flute''}
         \label{fig:uni-flute}
     \end{subfigure}
     \hfill
     \begin{subfigure}[b]{0.49\textwidth}
         \centering
         \includegraphics[width=\textwidth]{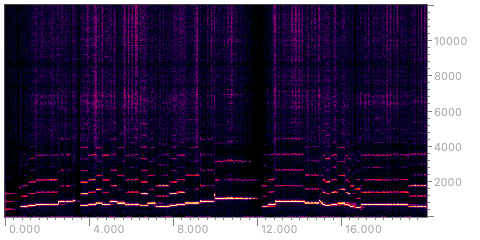}
         \caption{10-step sampling with our proposed angle schedule for text prompt: ``flute''}
         \label{fig:lin-flute}
     \end{subfigure}
     \begin{subfigure}[b]{0.49\textwidth}
         \centering
         \includegraphics[width=\textwidth]{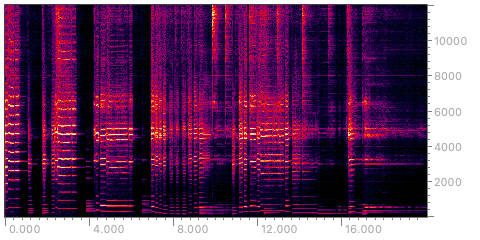}
         \caption{10-step sampling with uniform angle schedule for text prompt: ``saxophone''}
         \label{fig:uni-saxophone}
     \end{subfigure}
     \hfill
     \begin{subfigure}[b]{0.49\textwidth}
         \centering
         \includegraphics[width=\textwidth]{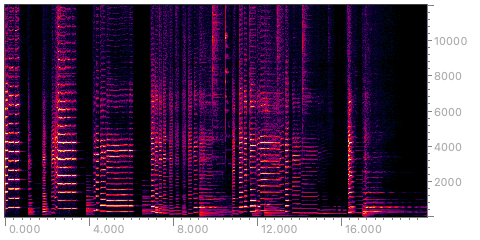}
         \caption{10-step sampling with our proposed angle schedule for text prompt: ``saxophone''}
         \label{fig:lin-saxophone}
     \end{subfigure}
     \begin{subfigure}[b]{0.49\textwidth}
         \centering
         \includegraphics[width=\textwidth]{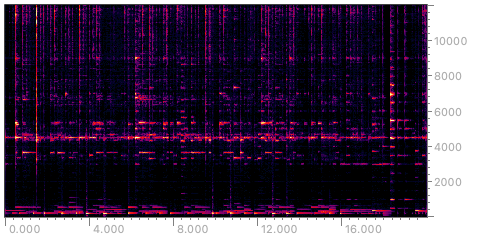}
         \caption{10-step sampling with uniform angle schedule for text prompt: ``acoustic guitar''}
         \label{fig:uni-acoustic}
     \end{subfigure}
     \hfill
     \begin{subfigure}[b]{0.49\textwidth}
         \centering
         \includegraphics[width=\textwidth]{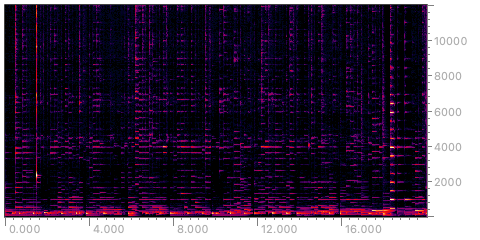}
         \caption{10-step sampling with our proposed angle schedule for text prompt: ``acoustic guitar''}
         \label{fig:lin-acoustic}
     \end{subfigure}
     \caption{Spectrograms of generated samples with uniform (left) and our proposed (right) angle schedules}
     \label{fig:spectrograms}
\end{figure}

We further qualitatively analyze the quality of the generated samples using some simple text prompts of instruments, i.e., flute, saxophone, and acoustic guitar, by pair-wise comparing their spectrograms as illustrated in Figure \ref{fig:spectrograms}. In the case of ``flute'', sampling with the proposed angle schedule results in a piece of naturally sound music, being more saturated in high-frequency bands and even remedying the breathiness of flute playing, as shown in Figure \ref{fig:lin-flute}. On the contrary, we can observe from the spectrogram in Figure \ref{fig:uni-flute} that the sample generated with a uniform angle schedule is comparatively monotonous. In the case of ``saxophone'', the uniform angle schedule leads to metallic sounds that are dissonant, as revealed by the higher energy in 3kHz to 6kHz frequency bands shown in Figure \ref{fig:uni-saxophone}. In comparison, the frequency bands are more consistent in Figure \ref{fig:lin-saxophone}, when the proposed schedule is used. While the comparatively poorer sample quality using the uniform schedule could be caused by the limited number of sampling steps, we also show the spectrograms after increasing the sampling steps from 10 to 20. In the case of ``acoustic guitar'', when taking 20 sampling steps, the samples generated with both angle schedules sound more natural. However, in Figure \ref{fig:uni-acoustic}, we witness a horizontal line around the 4.4kHz frequency band, which is unpleasant to hear. Whereas, the sample generated by our proposed schedule escaped such an acoustic issue, as presented in Figure \ref{fig:lin-acoustic}.

\begin{table}[h]
  \caption{The objective measures for the ablation study on architectures.}
  \label{tab:ablation-architecture}
  \centering
  \begin{tabular}{ccc}
    \toprule
    % \multicolumn{2}{c}{Part}                   \\
    % \cmidrule(r){1-2}
    \textbf{Architecture}  & \textbf{Velocity MSE} ($\downarrow$) & \textbf{SI-SNRi} ($\uparrow$) \\
    \midrule
    UNet-1D \cite{schneider2023mo}  & 0.13 & 5.33 \\
    UNet-2D \cite{riffusion}  & 0.15 & 4.96 \\
    \midrule
    \textbf{DPD} (Ours) & \textbf{0.12} & \textbf{6.15} \\
    \bottomrule
  \end{tabular}
\end{table}

\section{Ablation Study on Architectures}
To examine the superiority of our proposed dual-path model in Figure \ref{fig:dualpath-diff}, we also study the ablation of network architectures. In particular, to focus on the denoising capability of different architectures, we only take a subset of the training data (approximately 5k hours of music data) to train different networks with the same optimization configurations -- 100k training steps using AdamW optimizer with a learning rate of $5\times 10^{-4}$ and a batch size of $96$ on 8 NVIDIA V100 GPUs. For a fair comparison, we train the UNet-1D\footnote{Our implementation of UNet-1D relied on https://github.com/archinetai/a-unet.} and the UNet-2D\footnote{Our implementation of UNet-2D relied on https://huggingface.co/riffusion/riffusion-model-v1.} with comparable numbers of parameters (approximately 300M). Note that the FAD and MCC measures are not suitable for evaluating the performance of each forward pass of the trained network for denoising. In addition to the training objective, i.e., the velocity MSE, we use the scale-invariant signal-to-noise ratio (SNR) improvements (SI-SNRi) \cite{luo2020dual, lam2021effective} between the true latent $\mathbf{z}$ and the predicted latent $\hat{\mathbf{z}}=\alpha_\delta \mathbf{z}_\delta-\sigma_\delta\hat{\mathbf{v}}_\theta(\mathbf{z}_\delta;\mathbf{c})$. The results are shown in Table \ref{tab:ablation-architecture}, where our proposed dual-path architecture outperforms the other two widely used UNet-style architectures in terms of both the velocity MSE and SI-SNRi.

\section{Qualitative Evaluation}
\begin{figure}[h]
     \centering
     \includegraphics[width=0.68\textwidth]{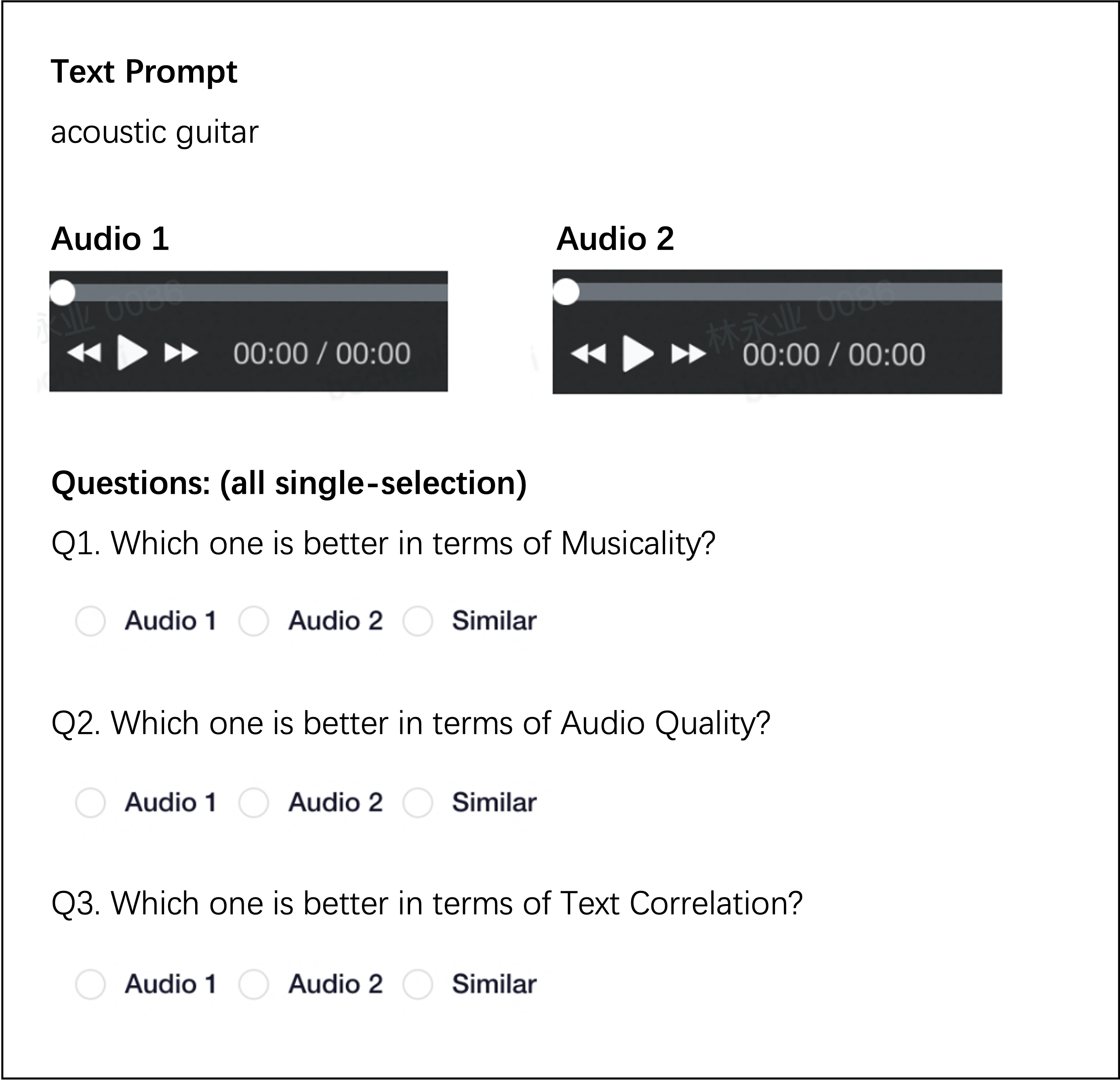}
     \caption{The user interface for music producers in each pair-wise comparison}
     \label{fig:UI}
\end{figure}

To conduct a pair-wise comparison, each music producer is asked to fill in the form composed of three questions. Specifically, we present the user interface for each pair-wise comparison in Figure \ref{fig:UI}.

%% file: algo.tex
\algrenewcommand\algorithmicindent{0.5em}%
\begin{figure*}[t]
\centering
\begin{minipage}[t]{0.49\textwidth}
\setcounter{algorithm}{0}
\begin{algorithm}[H]
  \caption{Music Generation} \label{alg:music-gen}
  \small
  \begin{algorithmic}[1]
    \State \textbf{given} $\mathcal{D}_\phi$, $\hat{\mathbf{v}}_\theta$, $T$, $\omega_1,\dotsc,\omega_T$
    \State \textbf{input} Music/text prompt $\mathcal{P}$\\
    \State Initialize $\delta_T=\pi/2$
    \State Compute the MuLan tokens for $\mathcal{P}$: $\mathbf{c}_{1:T_\text{cnd}}$
    \State Generate $\mathbf{u}_{1:T_\text{ST}}$ from $\mathbf{c}_{1:T_\text{cnd}}$ with LM \Comment{(\ref{eq:autoregressive-modeling})}
    \State Sample $\mathbf{z}_{\delta_T} \sim \mathcal{N}(\mathbf{0}, \mathbf{I})$
    \For{$t=T$ to $1$}
      \State Prepare condition: $\mathbf{c}=\{\mathbf{u}_{1:T_\text{ST}}, {[\delta_t]}_{r=1}^{L}\}$ \Comment{(\ref{eq:condition})}
      \State Update angle: $\delta_{t-1}=\delta_{t}-\omega_{t}$ \Comment{(\ref{eq:delta-t-1})}
      \State $\mathbf{z}_{\delta_{t-1}} =\cos({\omega_t})\mathbf{z}_{\delta_t}-\sin({\omega_t})\hat{\mathbf{v}}_\theta(\mathbf{z}_{\delta_t};\mathbf{c})$ \Comment{(\ref{eq:update-rule})}
    \EndFor
    \Repeat
        \State \textbf{pass} $\mathbf{c}_{1:T_\text{cnd}}$, $\mathbf{u}_{1:T_\text{ST}}$ and $\mathbf{z}_0$ to \textbf{Algorithm 2}
    \Until $\mathbf{z}_0$ reaches the desired length
    \State \textbf{return} $\mathcal{D}_\phi(\mathbf{z}_0)$
  \end{algorithmic}
\end{algorithm}
\end{minipage}
\hfill
\begin{minipage}[t]{0.49\textwidth}
\setcounter{algorithm}{1}
\begin{algorithm}[H]
% \begin{algorithm}[t]
  \caption{Music Continuation} \label{alg:music-continue}
  \small
  \begin{algorithmic}[1]
    \State \textbf{given} $\mathcal{D}_\phi$, $\hat{\mathbf{v}}_\theta$, $T$, $M$, $\omega_1,\dotsc,\omega_T$
    \State \textbf{input} Music $\mathbf{z}_0$ and $\mathbf{c}_{1:T_\text{cnd}}$, $\mathbf{u}_{1:T_\text{ST}}$ (if provided)\\
    \State Denote $M_\text{ST}=\lceil T_\text{ST}/M\rceil$, $L_M=\lceil L/M\rceil$
    \State Initialize $\delta_T=\pi/2$
    \State Generate $\mathbf{u}_{T_\text{ST}:T_\text{ST}+M_\text{ST}}$ from $\mathbf{c}_{1:T_\text{cnd}}\oplus \mathbf{u}_{M_\text{ST}:T_\text{ST}}$
    \State Sample $\mathbf{z}_\text{new} \sim \mathcal{N}(\mathbf{0}, \mathbf{I}) \in\mathbb{R}^{L_M}$
    \State Save first chunk: $\mathbf{z}_\text{save}=\mathbf{z}_0[:L_M]$
    \State $\mathbf{z}_{\delta_T}=\mathbf{z}_0[L_M:]\oplus \mathbf{z}_\text{new}$
    \For{$t=T$ to $1$}
      \State Update $\boldsymbol\delta_\text{new}=[0]_{r=1}^{L-L_M}\oplus[\delta_t]_{r=1}^{L_M}$
      \State Prepare condition: $\mathbf{c}=\{\mathbf{u}_{M_\text{ST}:T_\text{ST}+M_\text{ST}}, \boldsymbol\delta_\text{new}\}$
      \State Update angle: $\delta_{t-1}=\delta_{t}-\omega_{t}$
      \State $\mathbf{z}_{\delta_{t-1}} =\cos({\omega_t})\mathbf{z}_{\delta_t}-\sin({\omega_t})\hat{\mathbf{v}}_\theta(\mathbf{z}_{\delta_t};\mathbf{c})$ 
    \EndFor
    \State \textbf{return} $\mathbf{z}_\text{save}\oplus\mathbf{z}_0$
  \end{algorithmic}
\end{algorithm}
\end{minipage}
\end{figure*}